%% file: main.tex
\documentclass[11pt,twoside]{article} % For LaTeX2e

\input{math_commands.tex}

\usepackage[utf8]{inputenc} % allow utf-8 input
\usepackage[T1]{fontenc}    % use 8-bit T1 fonts
\usepackage{booktabs}       % professional-quality tables
\usepackage{amsfonts}       % blackboard math symbols
\usepackage{nicefrac}       % compact symbols for 1/2, etc.
\usepackage{microtype}      % microtypography

\usepackage{subfigure}
\usepackage{epsf}
\usepackage{epsfig}
\usepackage{fancyhdr}
\usepackage{graphics}
\usepackage{graphicx}
\usepackage{psfrag}
\usepackage{fullpage}
\usepackage{pdfpages}
\usepackage{natbib}
\usepackage{url}% for url's in bib
\usepackage[colorlinks,linkcolor=magenta,citecolor=blue, pagebackref=true]{hyperref}
\renewcommand*{\backrefalt}[4]{%
    \ifcase #1  {(Not cited.)}%
    \or         {(Cited on page~#2.)}%
    \else       {(Cited on pages~#2.)}%
    \fi}
\usepackage{color}

\usepackage{amsthm}
\usepackage{amsmath}
\usepackage{amssymb,bbm}
\usepackage{caption}
\usepackage{algorithm}
\usepackage{textcomp}
\usepackage{siunitx}
\usepackage{wrapfig}
\usepackage{multirow}
\usepackage{multicol}% colors
\usepackage{caption}
\usepackage{algorithm}
\usepackage{algpseudocode}
\usepackage{amssymb}
\usepackage{graphicx}                        % \includegraphics
\usepackage{tikz}
\definecolor{softblue}{RGB}{180,185,230}
\definecolor{softgreen}{RGB}{190,235,190}
\definecolor{softred}{RGB}{240,180,170}
\definecolor{boxgray}{RGB}{235,235,235}% MNIST figure (tikzpicture)
\usetikzlibrary{positioning, arrows.meta}    % above=, >=Latex arrows
\usepackage{booktabs} 
\usepackage{multirow}  
\usepackage{graphicx}  
\usetikzlibrary{fit}

\newtheorem{assumption}{Assumption}
\def\DD{\mathbb{D}}

\usepackage{graphicx}
\usepackage{booktabs}
\renewcommand{\th}{\theta}
\newcommand{\tht}{\widetilde{\theta}}
\newcommand{\betat}{\widetilde{\beta}}
\newcommand{\ths}{\bm{\theta^\star}}
\newcommand{\ts}{t^\star}
\newcommand{\betas}{\bm{\beta^\star}}

\newcommand{\thb}{\bm{\theta}}
\newcommand{\betab}{\bm{\beta}}

\newcommand{\phat}{\widehat{p}}
\newcommand{\Shat}{\widehat{S}}
\newcommand{\Shatb}{\bm{\widehat{S}}}

\newcommand{\yb}{\bm{y}}

\newcommand{\xb}{\bm{x}}
\newcommand{\bb}{\bm{b}}

\newcommand{\zb}{\bm{z}}

\def\bSig\mathbf{\Sigma}

\newcommand{\tr}{{\text{tr}}}
\newcommand{\te}{{\text{te}}}

\newcommand{\II}{\mathcal{I}}
\newcommand{\LL}{\mathcal{L}}
\newcommand{\OO}{\mathcal{O}}
\newcommand{\Ibb}{\mathbb{I}}
\renewcommand{\Re}{\mathbb{R}}
\newcommand{\nbar}{\bar{n}}

\newcommand{\DP}{\text{DP}}
\begin{document}

\begin{center}

{\bf{\LARGE{Generalized Bayesian Clustering with Regression for Unaligned Longitudinal Binary Data
}}}
  
\vspace*{.2in}
{\large{
\begin{tabular}{c}
Khai Nguyen$^{1}$,
  Elizabeth Juarez-Colunga$^2$,
  Peter Mueller$^{3,4}$ 
\end{tabular}
}}

\vspace*{.2in}

\begin{tabular}{c}
$^1$Department of Statistics, Texas A\&M University
 \\
$^2$Department of Biostatistics and Informatics, University of
Colorado Anschutz
\\
$^3$Department of Statistics and Data Sciences, University of Texas at Austin \\
$^4$Department of Mathematics,
University of Texas at Austin
\end{tabular}

\vspace*{.2in}
\today

\vspace*{.2in}

\begin{abstract}
We propose a generalized Bayesian clustering with regression model for
unaligned longitudinal binary outcomes, motivated by seizure diary
data from the Human Epilepsy Project.  Seizure diaries are sparse,
irregularly observed, and vary enormously across patients.
A single fully-specified generative model tends to be either
misspecified or computationally inefficient. We address the challenge
by two strategies. We set up a regression by way of clustering as
model-based clustering using a mixture model. For the latter, we take a
generalized Bayesian perspective which replaces the full likelihood
with a loss-based update using a generalized likelihood. We combine a
trajectory similarity loss and a regression loss, so that clustering
is informed by both trajectory similarity and the prediction of
outcomes The trajectory similarity loss is constructed by representing
each trajectory as an (empirical) distribution of subsequences, called
\textit{reads}, and then is defined based on the sliced Wasserstein
distance between these empirical distributions.
This loss allows alignment-free comparison of sequences that are
irregularly observed or temporally misaligned, and it scales
quasi-linearly in trajectory length.
The regression loss is the negative log-likelihood of a probit regression.
A prior on the cluster-specific parameters is defined by way of a
Dirichlet process prior on the mixing measure.
\end{abstract}

\end{center}

\section{Introduction}
\label{section:introduction}

We introduce a novel model for regression-based clustering of
unaligned longitudinal binary data. The proposed model relies on a
generalized Bayesian paradigm~\citep{bissiri2016general}, avoiding the need for a detailed generative model. 
Clustering is set up using a two-component loss, including a
trajectory similarity loss and a regression loss. The trajectory
similarity loss involves a cluster-specific average trajectory, which can be interpreted as the trajectory of a typical patient. This provides an easily interpretable and easy to communicate
representation of clusters, replacing abstract parameters which often remain meaningless for clinical collaborators. The trajectory
similarity loss is based on sliced Wasserstein (SW)
distance~\citep{rabin2012wasserstein,nguyen2025introduction} between
distributions over ``reads" which are defined as subsequences of the
trajectory. The regression loss arises from a probit regression of binary
outcomes on lagged outcomes, treatment assignment, and baseline
covariates. A joint prior for parameters
indexing both components of the loss function
is defined by way of a nonparametric prior on a discrete mixing measure.
Posterior inference under the model remains computationally efficient
and tractable.  In experiments, the model recovers distinct and
homogeneous clusters, and allows for accurate prediction.  Inference under the
fitted model includes in particular recommendations for 
treatment-change.

In various clinical trial settings, unaligned longitudinal binary
data~\citep{diggle2002analysis, fitzmaurice2012applied} are gathered
through monitoring patient responses over time at irregular
intervals. The primary challenge involved in such datasets is the high
level of heterogeneity across patients, including in particular widely
varying treatment responses~\citep{varadhan2013estimation,
  kent2018personalized}.  Clustering based on the trajectories of
patient responses has a high level of clinical utility.  Recognizing
homogeneous subgroups of patients can aid in deciding treatment
choices for a particular patient.  However, clustering alone is
insufficient for predicting outcomes.  Meaningful prediction should
explicitly (beyond cluster membership) account for treatment and
covariates.  Clustering and regression complement each other since
clustering alone cannot predict the future response using the
information about cluster, and regression alone disregards the
heterogeneity of patient populations.

The proposed inference framework is motivated by the analysis of
seizure data from the Human Epilepsy Project
(HEP)~\citep{French_HumanEpilepsyProject, pellinen2020focal} (more
detail in the upcoming section).  Newly diagnosed patients are known
to have an increased chance of secondary seizures after every primary
event~\citep{bauman2021seizure, jafarpour2019seizure}, with
significant consequences for health and quality of life. At the same
time, epilepsy manifests with high variability both in frequency and
regularity of seizure events~\citep{chiang2020individualizing,
  haut2006seizure}.
However, this heterogeneity is difficult to quantify due to the
nature of the seizure diary data that is available in the HEP
study. The responses differ, and the timing of participation and study
length vary greatly across participants.  The recording process is
also subject to a high level of
missingness~\citep{miller2024long}. Reported seizures tend to be
underreported, since patients often fail to remember or recognize an
episode~\citep{fisher2012seizure, hoppe2007epilepsy}. Overall, the
dataset consists of non-aligned sequences of observations of highly
variable length and levels of missingness across patients. 

Clustering of longitudinal trajectories is not a new problem in
statistics and machine learning. Mixture models and hidden Markov
models have long been used to find subgroups that share similar
temporal
patterns~\citep{nagin1999analyzing,amato2025mmm,cantoni2025borrowing}. Joint
models that combine clustering with regression can borrow strength
across individuals in the same cluster. However, they are usually tied
to a specific generative model, which is problematic with highly
complex data formats like in the HEP study.  Generalized Bayesian
inference~\citep{bissiri2016general} addresses this by replacing the
likelihood of a generative model by a loss-based update, which buys
robustness to misspecification.  However, to the best of our
knowledge, there is no existing generalized Bayesian clustering and
regression model for this type of data. We address the gap for
coherent posterior inference in this setting by proposing to combine
two losses that represent trajectory similarity and prediction
accuracy to construct a generalized likelihood. 

For the construction of the trajectory similarity loss,  a core
challenge is the comparison of trajectories that are unevenly observed
and misaligned. Distances such as Euclidean distance assume
observations at aligned times, which is not usually feasible in
clinical data, especially not in outpatient data like the HEP
study. Dynamic Time Warping~\citep{berndt1994using} and other
alignment distances alleviate this assumption but do not scale well
with the length of the trajectory. In this paper, we develop an
alternative approach whereby each observation of the longitudinal data
is split into many subsequences that we call ``reads", that is, the
trajectory is replaced by a distribution of reads. Temporal dependence
is captured by using subsequent and overlapping reads. The main
feature of this approach is that it allows for the computation of
distances based on distances between distributions of reads, e.g., the
Wasserstein distance~\citep{peyre2020computational}. The main
advantage being that this process can proceed without
alignment. However,  Wasserstein distance is computationally
expensive~\citep{peyre2020computational} (super-cubic in the number of
reads). Instead we use the SW
distance, which is quasi-linear in the number of reads. To deal with
missing data, they are recorded as a neutral value rather than
dropped. As a result, a read spanning a gap in reporting still carries
information about the gap itself, and patients who share similar
missingness patterns end up with similar reads and are compared
accordingly.

For the regression loss, we keep the construction simple by basing it on the
negative log-likelihood of an autoregressive probit model. The
covariates in the probit regression include the lagged outcome
history, current and recent treatment assignments, and baseline
patient covariates.  We only evaluate the loss at observed time
points, treating missingness as ignorable under a missing-at-random
assumption (alternative choices are possible without changing the
overall framework and the upcoming discussion).  For the lagged
outcomes in the predictor of the probit regression, we record missing
values as a neutral value, such that the covariate history remains
well defined. 

The remainder of the article is organized as follows. Section~\ref{section:data} describes the HEP project. We then introduce the proposed model in Section~\ref{section:model}, including technical details of constructing the trajectory similarity loss and the regression loss. Section~\ref{sec:posterior_inference} discusses posterior inference of the proposed model and how to to perform prediction and clustering. In Section~\ref{section:experiments} we present the results for the HEP data 

\section{Human Epilepsy Project Data}
\label{section:data}

We first review the dataset. As described in~\citep{kanaster2026mixed}, HEP~\citep{French_HumanEpilepsyProject} diagnosed 448 people with focal epilepsy at 34 clinical centers between 2012 and 2017. To qualify, participants had to be between 12 and 60 years old at diagnosis and had to enroll within four months of starting treatment. Daily seizure counts were logged through an electronic diary~\citep{fisher2010tracking} for three years of follow-up. Some individuals were dropped for missing diary entries, insufficient tracked days, or missing medication data, leaving 407 in the final cohort. Study eligibility, the diary process, and tracking protocol are described more fully elsewhere~\citep{miller2024long,pellinen2020focal}.

Women made up 60.4\% of the cohort, and the mean age at enrollment was 32.5 years (SD = 13.8). Levetiracetam was the most common starting medication (40.8\%). Sodium channel blockers other than Lamotrigine came next (20.4\%), followed by Lamotrigine (16.7\%) and combination therapy (15.0\%). Before treatment began, participants had a median of 8 seizures, at a median rate of about 2.0 per month. Participants tracked only 43.4\% of days on average (SD = 30.9\%) across a mean 24-month follow-up, and that ranged all the way from 0.15\% to 100\% depending on the person. Reported seizure rates per tracked month averaged 6.7 (SD = 19.4) but the median was just 0.17, with values as high as 187, seizure burden varies enormously between individuals. 
We refer to Table~\ref{tab:hep_characteristics} for more detail.  The hazard of a seizure event fell over the course of the study, dropping sharply in the first year, which fits with what is expected once treatment starts~\citep{kanaster2026mixed}. 

\begin{table}[htbp]
\centering
\caption{Characteristics of Human Epilepsy Project participants~\citep{kanaster2026mixed}, stratified by sex. ``Pre-Tx'' refers to the period between seizure onset and treatment initiation.}
\label{tab:hep_characteristics}
\scalebox{0.9}{
\begin{tabular}{lccc}
\toprule
 & Female (N=246) & Male (N=161) & Overall (N=407) \\
\midrule
Age (Years) & & & \\
\quad Mean (SD) & 32.1 (13.2) & 33.1 (14.6) & 32.5 (13.8) \\
\quad Median [Min, Max] & 31 [11, 60] & 33 [11, 64] & 32 [11, 64] \\
\addlinespace
Completed Higher Education & 128 (52.0\%) & 82 (50.9\%) & 210 (51.6\%) \\
Part-Time or Unemployed & 54 (22.0\%) & 27 (16.8\%) & 81 (19.9\%) \\
Abnormal Findings on MRI & 37 (15.0\%) & 29 (18.0\%) & 66 (16.2\%) \\
Injury Prior to Diagnosis & 109 (44.3\%) & 71 (44.1\%) & 180 (44.2\%) \\
Family History of Seizures & 76 (30.9\%) & 48 (29.8\%) & 124 (30.5\%) \\
\addlinespace
Anti-Seizure Medication & & & \\
\quad Combination Therapy & 34 (13.8\%) & 27 (16.8\%) & 61 (15.0\%) \\
\quad Levetiracetam & 97 (39.4\%) & 69 (42.9\%) & 166 (40.8\%) \\
\quad Lamotrigine & 50 (20.3\%) & 18 (11.2\%) & 68 (16.7\%) \\
\quad Other Sodium Channel Blocker & 50 (20.3\%) & 33 (20.5\%) & 83 (20.4\%) \\
\quad Other Anti-Seizure Medication & 8 (3.3\%) & 6 (3.7\%) & 14 (3.4\%) \\
\addlinespace
Duration of Pre-Tx Seizures (Months) & & & \\
\quad Mean (SD) & 36.1 (71.9) & 37.1 (76.6) & 36.5 (73.7) \\
\quad Median [Min, Max] & 7.46 [0.0329, 531] & 7.85 [0.0329, 612] & 7.62 [0.0329, 612] \\
\addlinespace
Total Number of Pre-Tx Seizures & & & \\
\quad Mean (SD) & 309 (2724) & 156 (620) & 249 (2153) \\
\quad Median [Min, Max] & 8 [1, 42123] & 7 [1, 6751] & 8 [1, 42123] \\
\addlinespace
Rate of Pre-Tx Seizures per Month & & & \\
\quad Mean (SD) & 10.8 (31.5) & 14.3 (47.7) & 12.2 (38.7) \\
\quad Median [Min, Max] & 2.08 [0.0186, 289] & 1.43 [0.0114, 396] & 1.96 [0.0114, 396] \\
\addlinespace
Duration of Follow-up (Days) & & & \\
\quad Mean (SD) & 730 (361) & 717 (373) & 725 (365) \\
\quad Median [Min, Max] & 830 [3, 1097] & 800 [7, 1096] & 816 [3, 1097] \\
\addlinespace
Total Number of Days Tracked & & & \\
\quad Mean (SD) & 327 (326) & 364 (350) & 342 (335) \\
\quad Median [Min, Max] & 206 [1, 1096] & 212 [1, 1096] & 207 [1, 1096] \\
\addlinespace
Percent of Days Tracked (\%) & & & \\
\quad Mean (SD) & 41.5 (30.3) & 46.3 (31.8) & 43.4 (30.9) \\
\quad Median [Min, Max] & 32.4 [0.154, 100] & 40.8 [0.186, 100] & 35.1 [0.154, 100] \\
\addlinespace
Total Number of Reported Seizures & & & \\
\quad Mean (SD) & 35.0 (96.4) & 25.2 (84.8) & 31.1 (92.0) \\
\quad Median [Min, Max] & 3 [0, 825] & 1 [0, 639] & 1 [0, 825] \\
\addlinespace
Rate of Seizures per Tracked Month & & & \\
\quad Mean (SD) & 6.82 (18.6) & 6.57 (20.6) & 6.72 (19.4) \\
\quad Median [Min, Max] & 0.407 [0, 187] & 0.0518 [0, 183] & 0.167 [0, 187] \\
\bottomrule
\end{tabular}
}
\vspace{4pt}
\begin{minipage}{\textwidth}
\footnotesize
Number (\%) of missing values: 2 (0.5\%) for Completed Higher Education, 5 (1.2\%) for Part-Time or Unemployed, 5 (1.2\%) for Abnormal Findings on MRI, 30 (7.4\%) for Injury Prior to Diagnosis, 17 (4.2\%) for Family History of Seizures, 15 (3.7\%) for Anti-Seizure Medication.
\end{minipage}
\end{table}

\section{Generalized Bayesian Clustering with Regression}
\label{section:model}
We introduce some notation by way of describing the data structure.
For each individual $i=1,\ldots,M$ (with $M>1$) we observe a  sequence
of seizure indicators $\yb_i=(y_{i1},\ldots,y_{iT_i})$, a treatment
sequence $\zb_i=(z_{i1},\ldots,z_{iT_i})$, and a vector of baseline
covariates $\bb_i$. Stacking across individuals gives the full data
$\yb=(\yb_1,\ldots,\yb_M)$, $\zb=(\zb_1,\ldots,\zb_M)$, and
$\bb=(\bb_1,\ldots,\bb_M)$. Here $T_i$ is the length of follow-up for
individual $i$, and it need not be the same across individuals. At
each time point $t=1,\ldots,T_i$, $y_{it}\in\mathcal{Y}=\{-1,1\}$
indicates whether a seizure occurred ($y_{it}=1$), 
$z_{it}\in \{0,1\}^D$ records the treatment status across $D$
distinct treatments at that time, and $\bb_i$ collects covariates measured
at baseline, such as demographic or clinical characteristics, and is
treated as fixed over the follow-up period. A key complication is that
$y_{it}$ is not observed at every time point for every
individual. Some days go untracked, so the seizure indicator is
missing. To handle this, let
$\OO_i\subseteq\{1,\ldots,T_i\}$ denote the set of time points at
which individual $i$'s seizure status was actually recorded. Then
$y_{it}$ is observed only for $t\in\OO_i$. For $t\notin\OO_i$, the
value is missing and must be accounted for separately. 

\subsection{A generalized mixture model for unaligned binary sequence data} 
\label{subsec:overview}
Model-based clustering is often set up by way of a mixture model.
For the desired clustering of patients in the HEP study this takes the following form
\begin{align}\label{mixture}
\mathcal{L}(\yb_i \mid G, \zb_i, \bb_i) \nonumber&= \sum_h \gamma_h\, \exp\left(-\ell(\yb_i, \zb_i, \bb_i ; \tilde \thb_h,\tilde\betab_h)\right) \\& = \int
\exp\left(-\ell(\yb_i, \zb_i, \bb_i  ; \thb_i,\betab_i)\right)\mathrm dG(\thb_i,\betab_i),
\end{align}
where $G= \sum_h \gamma_h
\delta_{(\tht_h,\betat_h)}$ is a discrete mixing measure
(writing $\ell(\ldots)$ in anticipation of the upcoming
discussion - see next).
Replacing the mixture by a model augmentation with latent
$(\th_i,\beta_i) \sim G$, sampling from the discrete mixing measure $G$,
introduces a partition as the configuration of ties among the
$(\th_i,\beta_i)$, $i=1,\ldots,n$.
In setting up the mixture model \eqref{mixture} we adopt a
\textit{generalized Bayesian} framework~\citep{bissiri2016general},
which replaces the likelihood of a detailed generative model with a
loss function $\ell( \yb, \zb, \bb\mid G)$.
Instead of requiring  a fully specified generative model for the data,
we update the prior belief and obtain a generalized  posterior as follows: 
\begin{align}\label{genLik}
    \pi(G \mid \yb, \zb, \bb) &\propto 
    \pi(G)\prod_{i=1}^M \mathcal{L}(\yb_i,  \mid G, \zb_i, \bb_i),
\end{align}
where $\pi(G)$ is the prior over $G$, the parameters of interest. This formulation offers two key advantages.
First, it does not require a  fully specified generative model. As a
result,  it is well-suited to settings where the data-generating
mechanism is complex, partially observed, or otherwise difficult to
characterize. This is the case with  longitudinal seizure trajectories
subject to substantial missingness and between-individual
heterogeneity. Second, generalized Bayes  
provides \textit{robustness} to model misspecification, as inference  is driven by a loss function rather than a likelihood and is therefore  less sensitive to departures from distributional assumptions.

For individual $i$, the individual-level loss jointly capturing the  trajectory similarity and regression objectives is
\begin{align}
    &\ell(\yb_i , \zb_i, \bb_i;  \betab_i, \thb_i) = 
    \alpha L_1(\yb_i ; \thb_i) +
     L_2(\yb_i ;  \zb_i, \bb_i, \betab_i),
\end{align}
where $\alpha> 0$ are hyperparameters, $ \{\betab_i, \thb_i\}_{i=1}^{M}$  denotes the parameters of interest across all individuals. $L_1(\yb_i \mid  \thb_i)$ is a \textit{trajectory similarity loss} 
measuring the fit of individual $i$'s outcome trajectory to the
cluster parameters $\thb_i$, and $L_2(\yb_i \mid  \zb_i, \bb_i,
\betab_i)$ is a \textit{regression loss} penalizing poor fit of
the binary outcome $y_{it} \in \{-1,1\}$ given treatments $\zb_i$,
baseline covariates $\bb_i$, and individual-level  coefficients
$\betab_i$. This compound loss structure enables the model to
simultaneously refine cluster assignments and improve regression
performance, borrowing strength across individuals  within the same
cluster. 

Back to the mixture model \eqref{mixture}. The mixture can
equivalently be written as a hierarchical model with latent
patient-specific variables
$(\thb_i,\betab_i) \sim G$. Using the generalized log likelihood
$\ell(\yb_i;\ldots)$ and adding a Ferguson-Dirichlet (DP) prior
\citep{ferguson1973bayesian} for $G$ we have the generalized hierarchical
model
\begin{align} \label{prior:DP}
    \ell(\yb_i , \zb_i, \bb_i;  \betab_i, \thb_i) & = 
    \alpha L_1(\yb_i ;  \thb_i) + 
     L_2(\yb_i ;  \zb_i, \bb_i, \betab_i),\\
    (\thb_i, \betab_i) \mid G & \sim G,  \nonumber \\
    G & \sim \DP(\alpha, P_0), \quad
    P_0 = P_0^{(1)} \otimes P_0^{(2)}, \nonumber
\end{align}
where $G$ is a random discrete measure drawn from a
Ferguson-Dirichlet process  with concentration parameter $\alpha > 0$ and base
measure $P_0$.  The base measure $P_0$ factorizes as a product of two
independent  measures $P_0^{(1)}$ and $P_0^{(2)}$, serving as the
respective  priors for $\thb_i$ and $\betab_i$. The discreteness of
$G$ induces clustering. In particular, individuals sharing
the same atom $(\thb_i, \betab_i)$ are assigned to the same cluster,
and the number of clusters $K$ is estimated from the data.
Let then $\{(\thb^*_1,\betab^*_1),\ldots,(\thb^*_K,\beta^*_K)\}$ denote the set
of unique $(\thb_i,\betab_i)$, with $(\thb^*_k,\betab^*_k)$ being the shared
unique values for all patients in cluster $k$.
By putting a DP prior
on $(\thb_i, \betab_i)$, inference of the random parition is based
on both regression loss and trajectory similarity loss.  

In the upcoming discussion we introduce a specification of $L_1$ using
sliced Wasserstein distance (SW) and of $L_2$ using a probit
regression model. We therefore refer to model
(\ref{genLik}-\ref{prior:DP}) as the DP-SW-Probit model.

% Crucially, the unique values $\{\thb^*_k\}_{k=1}^{K}$ of  $\thb_i$ play the role of \textit{typical patients}. Each $\thb^*_k$ represents the prototypical outcome trajectory shared by  all individuals in cluster $k$, capturing the dominant patterns of  treatment response present in the population. 

% \begin{align}
%      &\ell(\yb_i ; \zb_i, \bb_i, \betab_i, \thb_i)= \exp\left(-L_1(\yb_i \mid \thb_i) - L_2(\yb_i \mid \zb_i, \bb_i, \betab_i)  \right) \\
%         &(\thb_i, \betab_i) \mid G 
%         \sim G \\
%        & G 
%         \sim DP(\alpha, P_0)\\
%         &P_0 
%         =  P_0^{(1)}\otimes P_0^{(2)}
%     \end{align}

\subsection{Trajectory Similarity Loss}
\label{subsec:trajectory_similarity_loss}

The goal of the trajectory similarity loss is to quantify  if the
individual  $i$'s observed outcome trajectory $\yb_i$ is similar to
a cluster-specific trajectory $\thb_i$ of their assigned cluster (and
not using $\betas_k$ in $L_1$).
Specifically, we define
\begin{align}\label{L1}
  L_1(\yb_i \mid (\thb_i,\betab_i)=(\ths_k,\betas_k) )=
  \DD(\yb_i,\ths_k),
\end{align}
where $\DD(\cdot, \cdot)$ is a distance measure between two  time
series that does not require temporal alignment. This is  particularly
important in our setting, as individuals may have  different follow-up
lengths $T_i$, irregular observation times, making standard pointwise
distances such as the Euclidean distance inappropriate. By allowing
for unaligned comparisons, $\DD$ can capture structural  similarities
in outcome trajectories across individuals regardless of their
temporal irregularities. We also want $\DD$ to be computationally
efficient as the length of data points can be large.

\paragraph{From time series to sets of reads} To construct a
computational efficient distance  $\DD$, we first convert each time series into
a set of  overlapping consecutive subsequences, which we refer to as
\textit{reads}. The set of reads of size $H$ for  individual $i$ is
then defined as 
\begin{align}
    \mathcal{R}(\yb_i) = \left\{ \rb_{it} = (y_{it}, y_{i,t+1}, \ldots, y_{i,t+H-1}) \in \Re^H : t, t+H-1 \in \{1,\ldots,T_i\} 
    \right\},
\end{align}
where each read $\rb_{it} \in \Re^H$ is a consecutive
subsequence  of length $H$, capturing a local pattern of the outcome
trajectory  over a window of $H$ consecutive time points. This
representation  converts $\yb_i$ into a set of points in
$\Re^H$. By changing to sets, the representation is  invariant
to temporal alignment across  individuals since reads from two
different individuals need not  correspond to the same time points.
For $t \notin \OO_i$,  we record $y_{it} = 0$, reflecting a neutral
outcome  at unobserved time points. In addition, imputing missing
outcomes with a fixed value is simple and convenient, and has the
added benefit of allowing individuals with similar missingness
patterns to be grouped together.

\paragraph{Sliced Wasserstein distance} Given the set of reads $\mathcal{R}(\yb_i)$ and $\mathcal{R}(\thb_i)$,  we define $\DD$ as  SW distance
\citep{rabin2012wasserstein,nguyen2025introduction} between the corresponding empirical distributions of reads. Let  $\mu_{\yb_i} = \frac{1}{|\mathcal{R}(\yb_i)|}\sum_{\rb \in  \mathcal{R}(\yb_i)} \delta_{\rb}$ and $\mu_{\thb_i} =  \frac{1}{|\mathcal{R}(\thb_i)|}\sum_{\rb \in \mathcal{R}(\thb_i)}  \delta_{\rb}$ denote the empirical measures over the reads of  $\yb_i$ and $\thb_i$ respectively, the distance $\DD$ 
is then defined as follows:
\begin{align}
    \DD(\yb_i, \thb_i) = \text{SW}_p^p(\mu_{\yb_i}, \mu_{\thb_i}) 
    =\int_{\mathbb{S}^{H-1}} W_p^p\left(\mathcal{P}^\vartheta_\sharp 
    \mu_{\yb_i},\, \mathcal{P}^\vartheta_\sharp \mu_{\thb_i}\right) 
    \mathrm d\sigma(\vartheta),
\end{align}
where $\mathbb{S}^{H-1}$ is the unit sphere in $\Re^H$,  $\sigma$ is the uniform measure on the unit sphere $\mathbb{S}^{H-1}$, though other choices are possible~\citep{nguyen2023energy},  $\mathcal{P}^\vartheta_\sharp \mu_{\yb_i}$ denotes the pushforward of $\mu_{\yb_i}$ under the projection $\mathcal{P}^\vartheta(\rb) = \langle \rb, \vartheta \rangle$ onto direction $\vartheta$, and $W_p$ is the $p$-Wasserstein distance on $\Re$. For a definition of pushforward, let $(\mathcal{X}_1, \Sigma_1)$ and $(\mathcal{X}_2, \Sigma_2)$ be measurable spaces, and suppose $f : \mathcal{X}_1 \to \mathcal{X}_2$ is a measurable map. For a measure $\mu$ on $(\mathcal{X}_1, \Sigma_1)$, the push-forward measure $f \sharp \mu$ on $(\mathcal{X}_2, \Sigma_2)$ is given by
$
    f \sharp \mu(B) = \mu\left(f^{-1}(B)\right), \ \forall B \in \Sigma_2.
$ In practice, the integral over $\mathbb{S}^{H-1}$ is approximated by averaging over $L$ randomly sampled projection directions, yielding the Monte Carlo approximation~\citep{bonneel2015sliced,nguyen2024quasimonte}:
\begin{align}
    SW_p^p(\mu_{\yb_i}, \nu_{\thb_i}) \approx \frac{1}{L} 
    \sum_{\ell=1}^L W_p^p\left(\mathcal{P}^{\vartheta_\ell}_\sharp 
    \mu_{\yb_i},\, \mathcal{P}^{\vartheta_\ell}_\sharp 
    \nu_{\thb_i} \right)
\end{align}
where $\vartheta_1, \ldots, \vartheta_L \stackrel{\text{iid}}{\sim} 
\sigma$. SW   is particularly attractive in this setting for several reasons. First, it compares distributions of reads rather than aligned sequences, making it naturally robust to temporal misalignment and  differences in trajectory length. Second, each one-dimensional Wasserstein distance $W_p(\mathcal{P}^{\vartheta}_\sharp \mu_{\yb_i}, \mathcal{P}^{\vartheta}_\sharp \nu_{\thb_i})$ admits a closed-form solution
\begin{align}
    W_p^p\left(\mathcal{P}^{\vartheta}_\sharp \mu_{\yb_i},\mathcal{P}^{\vartheta}_\sharp \nu_{\thb_i}\right) 
    = \int_0^1 \left| F_{\mathcal{P}^{\vartheta}_\sharp \mu_{\yb_i}}^{-1}(u) 
    - F_{\mathcal{P}^{\vartheta}_\sharp \nu_{\thb_i}}^{-1}(u) \right|^p \mathrm du,
\end{align}
where $F_{\mathcal{P}^{\vartheta}_\sharp \mu_{\yb_i}}^{-1}$ and $F_{\mathcal{P}^{\vartheta}_\sharp \nu_{\thb_i}}^{-1}$ denote the quantile 
functions of the projected distributions. The quantile functions are
obtained in practice by sorting the projected empirical samples,
avoiding the need to solve a linear programming
problem~\citep{peyre2020computational}. As a result, $SW_p$  is highly
computationally efficient. Given $n_1 = |\mathcal{R}(\yb_i)|$ and $n_2
= |\mathcal{R}(\thb_i)|$ reads for the two time series, and $L$
projection directions, computing $SW_p(\mu_{\yb_i}, \nu_{\thb_i})$
requires: (i) projecting all reads onto each direction
$\vartheta_\ell$, at a cost of $\OO(L(n_1 + n_2)H)$, and (ii) evaluating
$L$ one-dimensional Wasserstein distances, each admitting a
closed-form solution via sorting at a cost of $\OO(L(n_1 \log n_1 +
n_2 \log n_2))$. The total time complexity for computing SW is
therefore $\OO\left(L(n_1 + n_2)    (H + \log(n_1 +
  n_2))\right)$, substantially more efficient than
Wasserstein distance in $\Re^H$, whose computation via linear
programming incurs a cost of $\OO((n_1 + n_2)^3 \log (n_1+n_2))$ in
general~\citep{peyre2020computational}. We recall that the number of
reads satisfies $n_i = T_i - H + 1$ after imputation, so the
complexity scales linearly in the trajectory length $T_i$ and the
number of projections $L$. 

\paragraph{Base measure} We specify the base measure 
\begin{align} \label{eq:base_measure} P_0^{(1)} =
\frac{1}{N}\sum_{j=1}^N \delta_{\boldsymbol \zeta_j}\end{align}

in the DP prior in \eqref{prior:DP} as a uniform
distribution over a set of $N$ candidate trajectories
$\{\boldsymbol \zeta_j\}_{j=1}^N$,  where each $\boldsymbol \zeta_j$ represents a possible
outcome trajectory.  The use of a finite discrete base measure, rather
than a continuous  one, is motivated by computational
considerations. In particular, a finite support  ensures fast mixing
of later Markov chain sampling as the  number of distinct atoms that
can be drawn from $G$ is bounded by  $N$. Moreover, it allows us to
precompute the distances $\DD$ before posterior simulation,
which is a significant speed-up given that evaluating $\DD$ is
often expensive. In our implementation, we use  $\{\boldsymbol \zeta_j\}_{j=1}^N =
\{\yb_j\}_{j=1}^M$ i.e., the pool of  candidate anchors is taken to be
the set of observed patient  trajectories themselves. This data-driven
choice is natural in our  setting since it ensures that every
$\thb_i$ is match with an actually observed
trajectory. Using actual observed trajectories, we could have direct clinical
interpretability to the cluster anchors, and avoids the need to
specify anchor trajectories a priori. 

\subsection{Regression Loss}

The regression loss is defined as the negative log-likelihood of a
\textit{probit regression} model for the binary outcome $y_{it}  
\in \{-1, 1\}$ with linear predictor:
\begin{align}
    \eta_{it} = \betab_i^\top \xb_{it}, \qquad \xb_{it} = \left(\yb_{i,t-H:t-1}^\top, \zb_{i,t-H:t}^\top, \bb_i^\top, 1\right)^\top,
\end{align}
where $\xb_{it} \in \Re^p$ is the design vector for individual
$i$ at time $t$, consisting of: the lagged outcome history
$\yb_{i,t-H:t-1} = (y_{i,t-H}, \ldots, y_{i,t-1})$ over the past $H$
time points, the treatment history $\zb_{i,t-H:t} = (z_{i,t-H},
\ldots, z_{i,t})$ over the same window,  baseline covariates $\bb_i$,
and an intercept term. For any missing entry at $t\notin
\OO_i$ appearing in the design vector $\xb_{it}$, we record
$y_{it} = 0$, similar to the imputation strategy adopted for the
trajectory similarity loss. This ensures that the design vector
$\xb_{it}$ is well-defined for all $i$ and $t$. 

The regression loss is then defined as the negative log-likelihood under
the probit model, evaluated over all time points 
with observed outcome:
\begin{align}\label{L2}
    L_2(\yb_i ; \zb_i, \bb_i, \betab_i) =  -\log \left(\prod_{t \in \OO_i}
    \Big[\Phi(\eta_{it})\Big]^{I(y_{it}=1)} \Big[1-\Phi(\eta_{it})\Big]^{I(y_{it}=-1)}\right),
\end{align} 
where $\Phi(\cdot)$ denotes the standard Normal cumulative
distribution function, so that $\Phi(\eta_{it}) = P(y_{it} = 1  \mid
\xb_{it}, \betab_i)$, and the product is taken only over  $t \in
\OO_i$. Missing observations are omitted from the  
likelihood. This is justified under the assumption that  data
are \textit{missing at random}, meaning that the probability of
missingness  depends only on the observed data and not on the
unobserved outcomes  themselves. Under the missing at random
assumption, the missing data mechanism is non-informative and can be
ignored during inference without inducing  bias. The base measure
$P_0^{(2)}$ for the individual-level coefficients  $\betab_i$ is
specified as a Gaussian prior, 
$
    P_0^{(2)} = \mathcal{N}(\mathbf{0}, \Sigma_0),
$
 where  $\Sigma_0$ is a prespecified covariance matrix encoding the prior  uncertainty on the regression coefficients.  Later, we simply choose $\Sigma_0=I$.

The hierarchical model (\ref{genLik}-\ref{prior:DP}) together with
loss \eqref{L1} and \eqref{L2} completes the construction of the
proposed DP-SW-Probit model.

%as the DP-SW-Probit model.

\section{Posterior inference}
\label{sec:posterior_inference}
\subsection{Posterior simulation}
\label{subsec:posterior_inference}
We implement posterior infrence by way of posterior Markov chain Monte
Carlo simluation, using the following transition probabilities.
We adopt Neal's Algorithm 8~\citep{neal2000markov}  to sample from the
generalized  posterior $\pi(\thb,\betab\mid \yb, \zb, \bb)$. Let
$S_i \in \{1, \dots, K\}$ denote the cluster assignment for patient
$i$, and let $\bm{\phi}_k^* = (\thb_k^*, \betab_k^*)$.
% denote the parameters
% associated with cluster $k$, we perform the following
% updates.
In the upcoming description of transition probabilities ``rest'' in
the conditioning set refers to all other currently imputed parameters
and the observed data.

\paragraph{Step 1: Latent variable augmentation} We use a
standard data augmentation scheme for probit
regression~\citep{chib1998analysis}. For each patient $i$ and observed
time  point $t \in \OO_i$, sample a latent probit score
$\lambda_{it}$  from a truncated normal distribution: 
\begin{align}
    \lambda_{it} \mid y_{it}, \eta_{it} \sim \text{TN}(\eta_{it}, 1\mid  
    a_{it}, b_{it}),
\end{align}
where $\eta_{it} = \betab_{S_i}^\top \xb_{it}$ and the truncation
bounds are: if $y_{it} = 0$: $(a_{it}, b_{it}) = (-\infty, 0)$, if
$y_{it} = 1$: $(a_{it}, b_{it}) = (0, \infty)$.
No probit scores are sampled for missing time points.

\paragraph{Step 2: Update cluster assignments ($S_i$)}  We follow
Neal's Algorithm 8 \citep{neal2000markov}. For each patient $i$
using $m$ potential values for a new cluster-specific parameters when
considering a new cluster, i.e., $S_i=K+1$ in Step 3, below.
\begin{enumerate}
    \item Denote $K^-$ as the number of distinct clusters currently occupied by patients other than $i$. If $i$ was the sole member of its cluster, discard that cluster.
    \item Draw $m$ (e.g., $m=1$) auxiliary parameters $\{\bm{\phi}_j\}_{j=K^-+1}^{K^-+m}$ independently from the base measure $P_0 = P_0^{(1)} \otimes 
    P_0^{(2)}$.
    \item Sample a new label $S_i$ from $\{1, \dots, K^-+m\}$ with probabilities:
    \begin{align}
        P(S_i = k \mid \text{rest}) \propto 
        \begin{cases} 
        \dfrac{n_{-i,k}}{M - 1 + \alpha} \, 
        \LL(\yb_i \mid \bm{\phi}_k^*) 
        & 1 \leq k \leq K^-, \\[10pt]
        \dfrac{\alpha/m}{M - 1 + \alpha} \, 
        \LL(\yb_i \mid  \bm{\phi}_k) 
        & K^- < k \leq K^-+m,
        \end{cases}
    \end{align}
    where $n_{-i,k}$ is the number of patients assigned to cluster $k$ excluding patient $i$, and the generalized likelihood $\LL(\yb_i \mid \bm{\phi}_k)$ is defined as
    \begin{align}
        \LL(\yb_i \mid \bm{\phi}_k) = 
        \exp\left(-\alpha \DD(\yb_i, \thb_k)\right) 
         \left(\prod_{t \in \OO_i} \mathcal{N}(\lambda_{it}; 
        \betab_k^\top \xb_{it}, 1)\right).
    \end{align}
\end{enumerate}

\paragraph{Step 3: Update cluster-specific parameters ($\bm{\phi}_k^*$)}  For each active cluster $k \in \{1, \dots, K\}$, we denote 
$\II_k = \{i : S_i = k\}$ as the set of patients assigned to cluster $k$. We update the unique cluster-specific parameters $\bm{\phi}_k^* = (\thb_k^*, 
\betab_k^*)$ as follows:

\subparagraph{Update $\betab_k^*$} Conditional on the latent
probit scores $\{\lambda_{it}\}$, the posterior for the unique  regression
coefficients $\betab_k^*$ is Gaussian by conjugacy with the
$\mathcal{N}(\mathbf{0}, \Sigma_0)$ prior: 
\begin{align}
    \betab_k^* \mid \text{rest} \sim \mathcal{N}\left(\bm{\mu}_{post}, 
    \Sigma_{post}\right),
\end{align}
where
$
    \Sigma_{post}^{-1}= \Sigma_0^{-1} + \sum_{i \in
      \II_k} \sum_{t \in \OO_i} \xb_{it}\xb_{it}^\top$
and    
    $\bm{\mu}_{post} = \Sigma_{post} \left(\sum_{i \in 
    \II_k} \sum_{t \in \OO_i} 
    \xb_{it}\lambda_{it}\right).
$

\subparagraph{Update $\thb_k^*$} The full conditional for the unique prototypical trajectory $\thb_k^*$ is
\begin{align}
    p(\thb_k^* \mid \text{rest}) \propto P_0^{(1)}(\thb_k^*) \exp\left(-\alpha \sum_{i \in \II_k} \DD(\yb_i, 
    \thb_k^*)\right).
\end{align}
 Since this distribution does not admit a closed form, we sample
 $\thb_k^*$ via a Metropolis-Hastings transition probability as
 follows.
 Propose a new candidate
 $\thb_k^{**} \sim P_0^{(1)}$, that is, draw uniformly at  random from the set of
 observed patient trajectories $\{\yb_1, \ldots, \yb_M\}$. The
 proposed candidate $\thb_k^{**}$ is accepted with probability 
\begin{align}
    \rho = \min\left(1, \frac{P_0^{(1)}(\thb_k^{**}) \exp\left(-\alpha \sum_{i \in \II_k} 
    \DD(\yb_i, \thb_k^{**})\right) P_0^{(1)}(\thb_k^*)}{P_0^{(1)}(\thb_k^*) 
    \exp\left(-\alpha \sum_{i \in \II_k} 
    \DD(\yb_i, \thb_k^*)\right) P_0^{(1)}(\thb_k^{**})}\right).
\end{align}
% Since $\tilde{P}_0^{(1)}$ assigns equal probability $1/M$ to each
% observed trajectory $\rho$ 
simplifying to
\begin{align}
    \rho = \min\left(1, \exp\left(-\alpha \sum_{i \in \II_k} 
    \left[\DD(\yb_i, \thb_k^{**}) - 
    \DD(\yb_i, \thb_k^*)\right]\right)\right).
\end{align}

Intuitively, the acceptance ratio compares the total dissimilarity of the proposed anchor $\thb_k^{**}$ to all patients currently in 
cluster $k$ against that of the current anchor $\thb_k^*$. If the proposed trajectory $\thb_k^{**}$ is on average closer to the trajectories of patients in $\II_k$. If $\sum_{i \in \II_k} \DD(\yb_i, \thb_k^{**}) < \sum_{i \in \II_k} \DD(\yb_i, \thb_k^*)$, then the exponent is positive and the proposal is accepted with  probability one. Conversely, if the proposed anchor is a worse representative of the cluster, the proposal is still accepted with a positive probability that decreases exponentially with the  increase in total dissimilarity, controlled by the scaling parameter $\alpha$. A larger $\alpha$ makes the sampler more selective, concentrating the anchor on trajectories that are genuinely close to all cluster members. A smaller $\alpha$ allows more exploratory moves across the pool of candidate anchors.

\subsection{Prediction and Clustering}
\label{subsec:prediction_and_clstering}

We discuss how to perform prediction and clustering under the proposed model.
For  already observed patients, our goal is to predict future outcomes,
averaging over imputed posterior cluster membership.
For new patients, the goal is to assign
cluster membership and predict future outcomes.
Since the generalized
Bayes framework does not include a notion of a joint distribution,
predictive distributions are not available for this purpose. We
therefore cast prediction and clustering of future observations
as thresholding and optimization - see next for details.

\paragraph{Prediction for Observed Patients} Given the cluster
assignments $\{S_i^{(b)}\}_{i=1}^M$ and estimated atoms
$\{\bm{\phi}_k^{*(b)}\}_{k=1}^K$ in each posterior Monte Carlo sample $b =
1,\ldots,B$, we predict the outcomes over the next $T'$ unobserved
time steps for each patient $i$ in the observed cohort ($T'=1$
recovers one-step-ahead prediction).
Considering prediction under a specific sequence of treatment assignments 
$\zb_{i,T_i+1:T_i+T'}$ over the forecast horizon
$\tau = 1,\ldots, T'$ and posterior sample $b$.
First we define a lagged outcome history as
\begin{align}
    \tilde{y}_{i,t}^{(b)} =
    \begin{cases}
        y_{it}               & t \in \OO_i,\\
        0                  & t \in \{1,\ldots,T_i\}\setminus\OO_i,\\
        \hat{y}_{i,t}^{(b)} & t > T_i,
    \end{cases}
\end{align}
using observed outcomes where available and imputing missing
ones as $0$. The design
vector for the prediction is then constructed as 
$ %\begin{align}
    \xb_{i,T_i+\tau}^{(b)} = \Bigl(
        \tilde{\yb}_{i,T_i+\tau-H:T_i+\tau-1}^{(b)\top},\;
        \zb_{i,T_i+\tau-H:T_i+\tau}^{\top},\;
        \bb_i^{\top},\; 1
     \Bigr)^{\top}.
$     % \end{align}

Future time steps are
replaced by previously predicted values from the same posterior
sample.  Within each sample $b$, the predicted outcome
$\tau$ steps ahead is recorded by the following thresholding, 
still conditional on parameters and cluster membership:
\begin{align} \label{phat}
    \hat{y}_{i,T_i+\tau}^{(b)} = \mathbb{I}\Big(
   \underbrace{\Phi\big({\betab_{S_i^{(b)}}^{*(b)}}^{\top}
            \xb_{i,T_i+\tau}^{(b)}\big)}_{\phat_{it}^{(b)}} >\delta
        \Big),
\end{align}
%  The posterior predictive probability, averaged over all $B$
% samples, is
% \begin{align}
%     \hat{p}_{i,T_i+\tau}
%         = \frac{1}{B}\sum_{b=1}^{B}
%           \Phi\left({\betab_{S_i^{(b)}}^{*(b)}}^{\top}
%           \xb_{i,T_i+\tau}^{(b)}\right).
% \end{align}
where $\delta>0$ is a threshold
($\delta=0.5$  in our implementation).
For later use we let $\phat_{it}= \Phi\left({\betab_{S_i}^{*}}^{\top}
            \xb_{i,T_i+\tau}\right)$ and  $\phat_{it}^{(b)}= \Phi\big({\betab_{S_i^{(b)}}^{*(b)}}^{\top}
            \xb_{i,T_i+\tau}^{(b)}\big)$ which is a  posterior sample of $\phat_{it}$.
Since the cluster assignment $S_i^{(b)}$ for observed
patients are already available in the current posterior sample,
no additional assignment to latent clusters 
is required. 

\paragraph{Clustering and Prediction for New Patients} For a new
patient $i^*$ with observed history $(\yb_{i^*}, \zb_{i^*},
\bb_{i^*})$, cluster assignment and prediction are
evaluated for each posterior sample $b$ (without re-running the full
posterior Markov chain Monte Carlo simulation).
As predicted cluster membership for the new patient we record
\begin{align}
    S_{i^*}^{(b)}
        = \arg\max_{k}\;
          n_k^{(b)}\,\LL\left(\yb_{i^*} \mid \bm{\phi}_k^{*(b)}\right),
\end{align}
where $n_k^{(b)} = |\II_k^{(b)}|$ is the size of cluster $k$
in posterior sample $b$, and the generalized likelihood is 
\begin{align}
    \LL\left(\yb_{i^*}\mid\bm{\phi}_k^{*(b)}\right)
        = \exp\left(-\alpha\,\DD(\yb_{i^*},\thb_k^{*(b)})\right)
          \left(\prod_{t\in\OO_{i^*}}
          \mathcal{N}\left(\lambda_{i^*t};\,
              {\betab_k^{*(b)}}^{\top}\xb_{i^*t},\,1\right)\right).
\end{align}
For each observed time point $t\in\OO_{i^*}$, the latent
probit score $\lambda_{i^*t}$ is sampled from
\begin{align}
    \lambda_{i^*t}\mid y_{i^*t},\eta_{i^*t}^{(b)}
        \sim \mathrm{TN}\left(\eta_{i^*t}^{(b)},1;\,a_{i^*t},b_{i^*t}\right),
\end{align}
where $\eta_{i^*t}^{(b)} = {\betab_k^{*(b)}}^{\top}\xb_{i^*t}$ and the
truncation bounds are as in  Step~1 of the Gibbs
sampler.
Note that $\lambda_{i^*t}$ is sampled separately for each candidate
cluster $k$ since the linear predictor depends on the cluster-specific
coefficients $\betab_k^{*(b)}$.

Autoregressive forecasting then proceeds identically to the observed patient case, replacing $S_i^{(b)}$ with $S_{i^*}^{(b)}$ and $\OO_i$ with
$\OO_{i^*}$ throughout. This procedure propagates two sources of uncertainty simultaneously i.e., uncertainty in the cluster-specific atoms $\bm{\phi}_k^{*(b)}$ and the
compounding autoregressive uncertainty that accumulates as predicted outcomes are fed back into the design vector over the forecast horizon.

\section{Results}
\label{section:experiments}
\subsection{Data coding}
Each patient visit is  characterised by a set of binary treatment
indicators that record concurrent anti-seizure medication (ASM).
For the HEP study cohort
\citep{pellinen2020focal},  we define five groups of ASM at
each time point $t$: (1) \textit{combination therapy}
(\texttt{combo}), indicating  $\geq 2$ concurrent ASMs,
(2) \textit{levetiracetam} (LEV), the  most commonly prescribed ASM in the
cohort (40.8\%), (3)  \textit{lamotrigine} (LTG, 16.7\%), (4)
\textit{other  sodium-channel blockers} (other-SCB, comprising
oxcarbazepine,  zonisamide, phenytoin, and eslicarbazepine, 20.4\%),
and  (5) \textit{other ASMs} (comprising topiramate, valproate,
brivaracetam, gabapentin, pregabalin, and perampanel, 3.4\%).
We define 5 binary treatment indicators corresponding to (1) through
(5). Including an increasing subset of these indicators defines
a nested sequence  of treatment-covariate scenarios $s_1 \subset s_2 \subset s_3 \subset s_4
\subset s_5$.
% where each scenario is a  strict superset of the previous.
In particular, $s_1$ uses an indicator for combination therapy only ($D =
1$ treatment covariates), $s_2$ adds and indicator for LEV ($D = 2$),
$s_3$  adds an indicator for LTG ($D = 3$), $s_4$ adds an indicator for
other-SCB ($D = 4$), and $s_5$ adds an indicator for other ASM ($D_s = 5$).  
% This nested  structure allows us to assess the incremental contribution  of each ASM group to the prediction of seizure outcomes.

Patient-level baseline  covariates are encoded as six binary
indicators and remain fixed across time for a given patient.
Sex is encoded as an indicator for female.
Age at enrollment is encoded by two indicators
$\Ibb[25 \leq \text{age}  \leq 39]$ and $\Ibb[40 \leq \text{age} \leq
64]$.
Educational  attainment is encoded as an indicator for ``higher
education'',
with high-school or below as the reference category (51.6\% of the
cohort completed higher education). Pre-treatment seizure burden is
encoded via two variables relative to a reference of $1$--$3$
seizures using indicators $\Ibb[4 \leq \text{seizures}_{\text{pre}} \leq 29]$ and $
\Ibb[\text{seizures}_{\text{pre}} \geq 30]$. Missing baseline
covariate values are imputed with the cohort mode
with a maximum of 30 (7.4\%)
missing values per covariate. Together these six indicators form the
baseline covariate vector $\bb_i \in \{0,1\}^{6}$. 

We compare inference under the proposed DP-SW-Probit model with 
inference under a Dirichlet process autoregressive probit model
(DP-Probit).
In more detail, DP-Probit is a special case of
DP-SW-Probit with $\alpha=0$, that is, without the
$L_1$ term in \eqref{prior:DP}, reducing
the DP-Probit to a conventional nonparametric Bayesian model with
a fully specified generative model.

Details of posterior inference, clustering, and
prediction are kept the same as discussed in 
Section~\ref{subsec:posterior_inference}--\ref{subsec:prediction_and_clstering}.
In  
the following experiments, we run 500 MCMC iterations with 100 burn-in
iterations and set the read size $H=7$.  For summarizing random
partition under the DP-SW-Probit and DP-Probit, we report the posterior
sample that minimizes the average variational of information
distance~\citep{meilua2007comparing,wade2018bayesian,dahl2022search}.
Below we refer to this posterior summary as the point estimate
$\Shatb=(\Shat_i;\; i=1,\ldots,n)$ of the random partition and the point estimate $(\hat{\thb}_k^*,\hat{\betab}_k^*)$ of the other parameters.

\subsection{Clustering of Patients and Treatments Recommendation}
\label{subsec:experiment_clustering}
{\em Clustering.}
We first discuss inference on partition of the study patients.
Table~\ref{tab:posterior_cluster_summary} reports
posterior expectation of the number of
clusters ($K$), posterior expectation and 95\% credible interval of 
average cluster size ($\nbar$),
as well as $K$ and $\nbar$ under a point estimate of the cluster
arrangement. 
For the DP-SW-Probit model, we report
results for $\alpha \in \{1, 10,100,1000\}$. 
From the table, we observe that adding the trajectory similarity loss
$L_1$ in the DP-SW-Probit model reduces $K$ compared to inference
under the DP-Probit without $L_1$ and
we find increasing $K$ with larger $\alpha$ under the DP-SW-Probit model.
In practice, we recommend to use the scale of
SW distances between observed data points as the reference for
choosing $\alpha$.  
We also observe that adding more covariates, from scenario $s_1$ to
scenario $s_5$, increases $K$.

% .
\begin{table}[!t]
\centering
\caption{Posterior summary of the estimated number of clusters $K$ and
  average cluster size $\nbar$ under DP-Probit and DP-SW-Probit (for
  $\alpha \in \{1,10,100,1000\}$), across treatment-covariate
  scenarios $s_1$--$s_5$. Columns report the posterior mean and 95\%
  credible interval for $K$ and $\nbar$, together with the corresponding
  values at the point-estimate $\widehat{S}$ of the partition.} 
\label{tab:posterior_cluster_summary}
\scalebox{0.8}{
\begin{tabular}{lllcccccc}
\toprule
Model & $\alpha$ & Case & $\mathbb{E}[K]$ & 95\% CI & $\mathbb{E}[\nbar]$ & 95\% CI & $K$ (point) & $\nbar$ (point) \\
\midrule
DP-Probit    & --    & $s_1$ & 26.09 & [26.00, 27.00] & 15.60 & [15.07, 15.65] & 26 & 15.65 \\
DP-SW-Probit & 1     & $s_1$ & 13.28 & [13.00, 14.00] & 30.69 & [29.07, 31.31] & 13 & 31.31 \\
DP-SW-Probit & 10    & $s_1$ & 14.55 & [14.00, 16.00] & 28.01 & [25.44, 29.07] & 15 & 27.13 \\
DP-SW-Probit & 100   & $s_1$ & 17.46 & [17.00, 19.00] & 23.33 & [21.42, 23.94] & 17 & 23.94 \\
DP-SW-Probit & 1000  & $s_1$ & 23.80 & [23.00, 26.00] & 17.12 & [15.65, 17.70] & 23 & 17.70 \\
\midrule
DP-Probit    & --    & $s_2$ & 25.15 & [25.00, 26.00] & 16.19 & [15.65, 16.28] & 25 & 16.28 \\
DP-SW-Probit & 1     & $s_2$ & 16.16 & [16.00, 17.00] & 25.20 & [23.94, 25.44] & 17 & 23.94 \\
DP-SW-Probit & 10    & $s_2$ & 17.88 & [17.00, 19.03] & 22.80 & [21.39, 23.94] & 18 & 22.61 \\
DP-SW-Probit & 100   & $s_2$ & 16.24 & [16.00, 17.00] & 25.08 & [23.94, 25.44] & 16 & 25.44 \\
DP-SW-Probit & 1000  & $s_2$ & 22.25 & [21.00, 23.00] & 18.33 & [17.70, 19.38] & 23 & 17.70 \\
\midrule
DP-Probit    & --    & $s_3$ & 28.03 & [28.00, 29.00] & 14.52 & [14.03, 14.54] & 28 & 14.54 \\
DP-SW-Probit & 1     & $s_3$ & 18.20 & [18.00, 19.00] & 22.37 & [21.42, 22.61] & 18 & 22.61 \\
DP-SW-Probit & 10    & $s_3$ & 19.27 & [19.00, 21.00] & 21.14 & [19.38, 21.42] & 19 & 21.42 \\
DP-SW-Probit & 100   & $s_3$ & 20.12 & [20.00, 21.00] & 20.24 & [19.38, 20.35] & 20 & 20.35 \\
DP-SW-Probit & 1000  & $s_3$ & 25.09 & [24.00, 26.00] & 16.23 & [15.65, 16.96] & 26 & 15.65 \\
\midrule
DP-Probit    & --    & $s_4$ & 35.15 & [35.00, 36.00] & 11.58 & [11.31, 11.63] & 35 & 11.63 \\
DP-SW-Probit & 1     & $s_4$ & 21.29 & [20.00, 23.00] & 19.14 & [17.70, 20.35] & 21 & 19.38 \\
DP-SW-Probit & 10    & $s_4$ & 24.11 & [24.00, 25.00] & 16.88 & [16.28, 16.96] & 24 & 16.96 \\
DP-SW-Probit & 100   & $s_4$ & 20.59 & [20.00, 21.03] & 19.79 & [19.36, 20.35] & 21 & 19.38 \\
DP-SW-Probit & 1000  & $s_4$ & 27.73 & [27.00, 28.00] & 14.68 & [14.54, 15.07] & 28 & 14.54 \\
\midrule
DP-Probit    & --    & $s_5$ & 25.01 & [25.00, 25.00] & 16.27 & [16.28, 16.28] & 25 & 16.28 \\
DP-SW-Probit & 1     & $s_5$ & 22.07 & [22.00, 23.00] & 18.44 & [17.70, 18.50] & 22 & 18.50 \\
DP-SW-Probit & 10    & $s_5$ & 21.18 & [20.00, 23.00] & 19.24 & [17.70, 20.35] & 21 & 19.38 \\
DP-SW-Probit & 100   & $s_5$ & 23.06 & [22.00, 24.00] & 17.65 & [16.96, 18.50] & 23 & 17.70 \\
DP-SW-Probit & 1000  & $s_5$ & 27.32 & [26.98, 28.00] & 14.90 & [14.54, 15.09] & 27 & 15.07 \\
\bottomrule
\end{tabular}
}
\end{table}

\begin{figure}[!t]
    \centering
    \begin{tabular}{c|c}
         \includegraphics[width=0.45\linewidth]{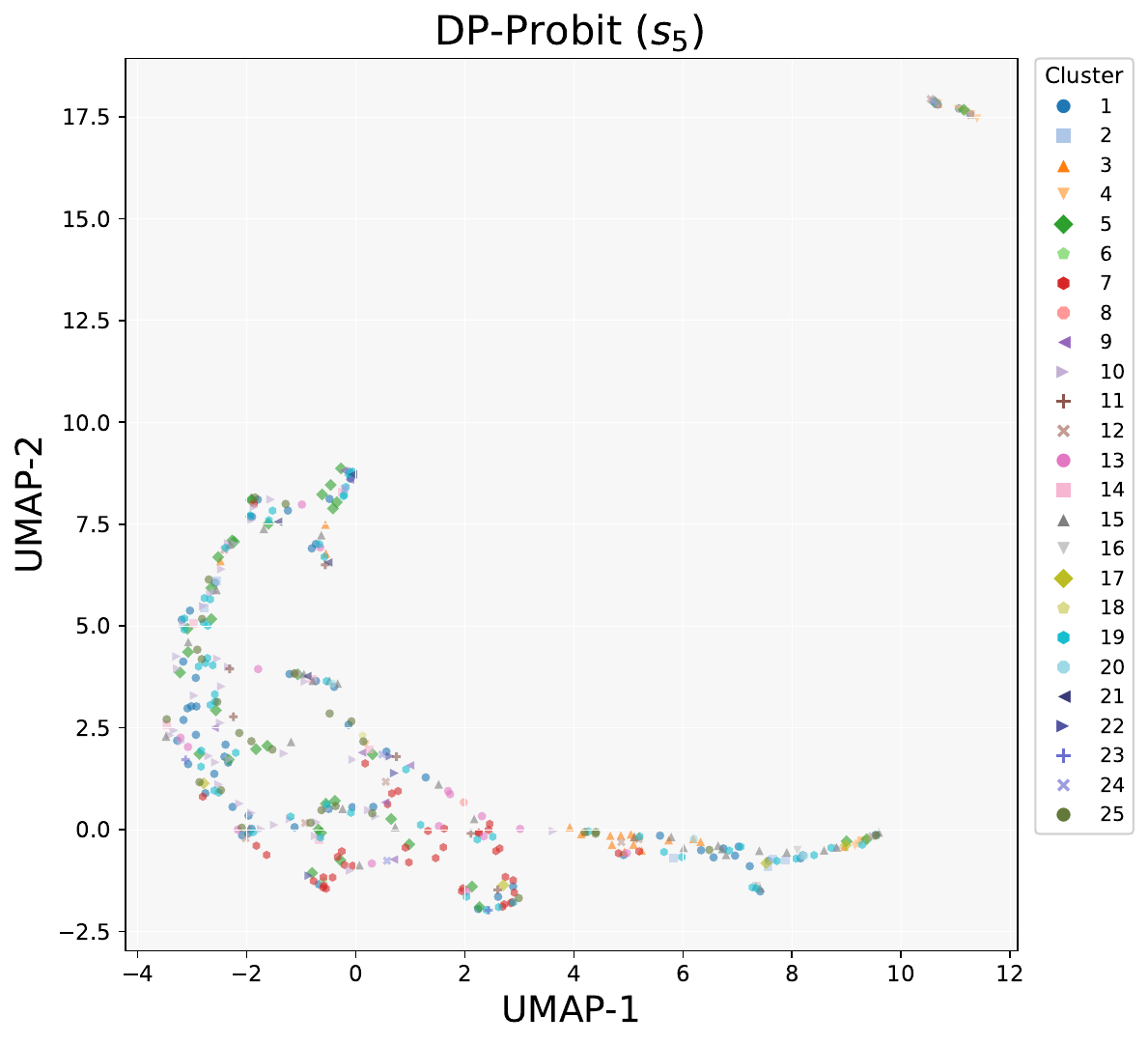}
         & \includegraphics[width=0.45\linewidth]{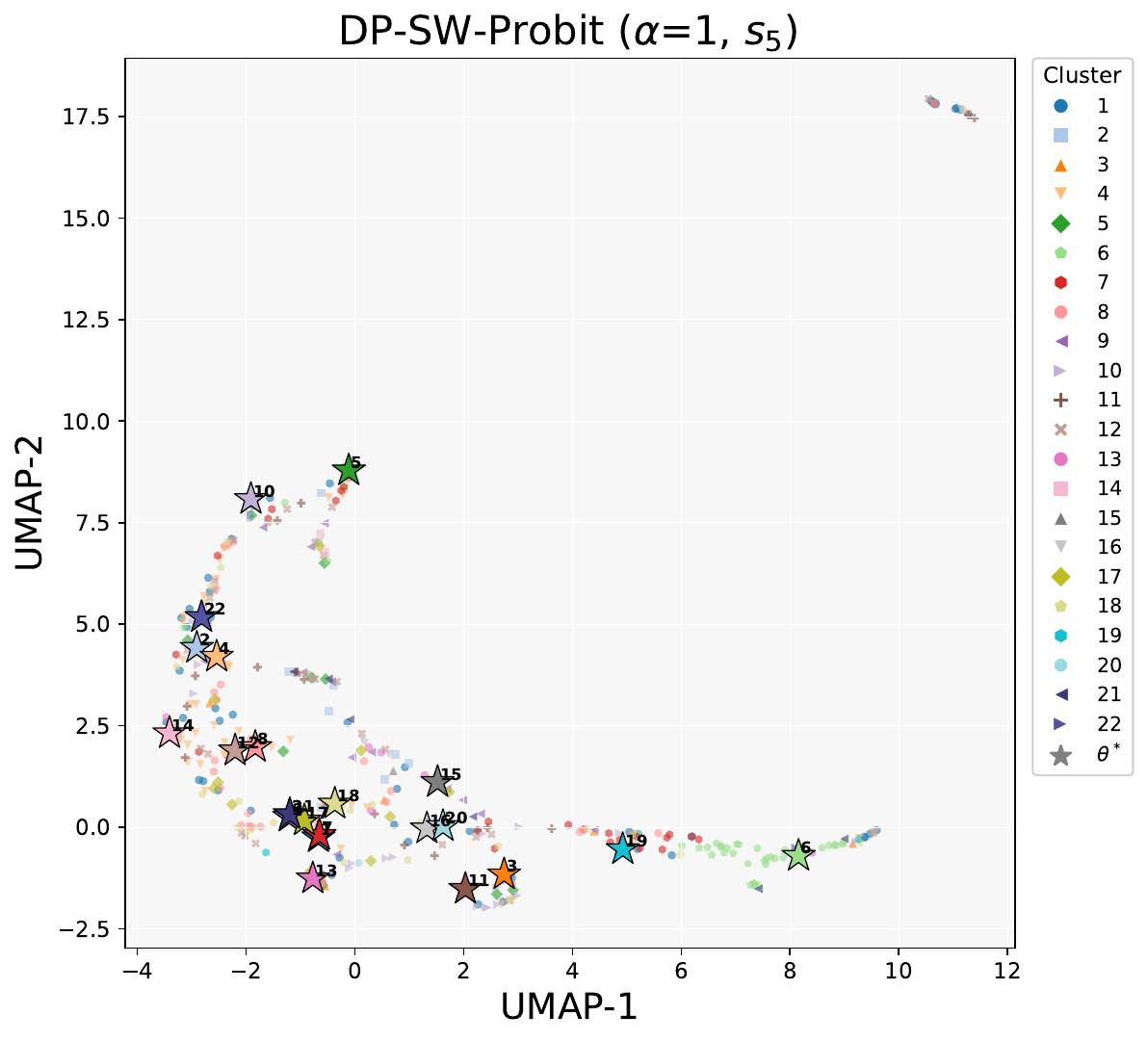} \\
         \\
         \includegraphics[width=0.45\linewidth]{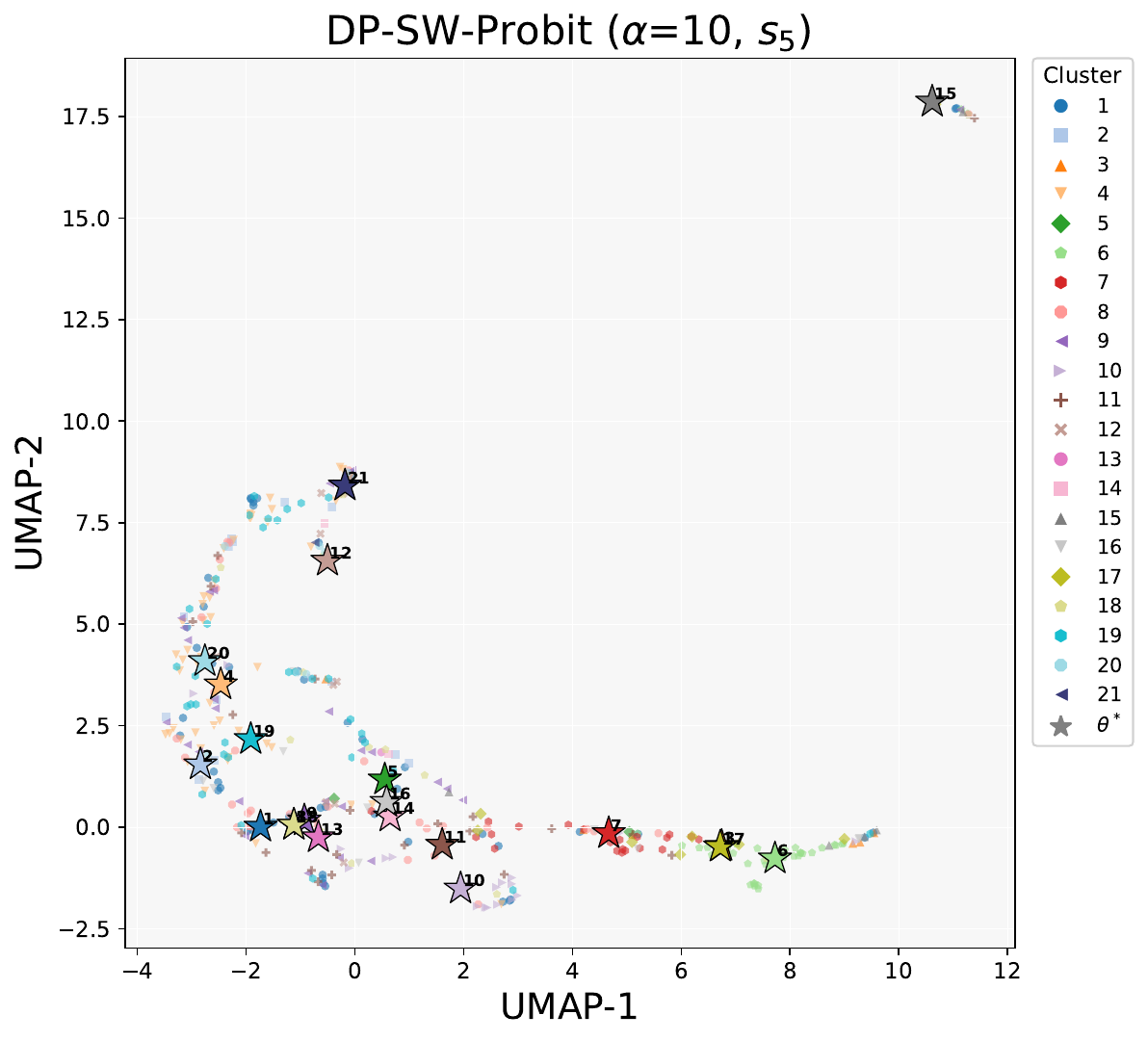}
         & \includegraphics[width=0.45\linewidth]{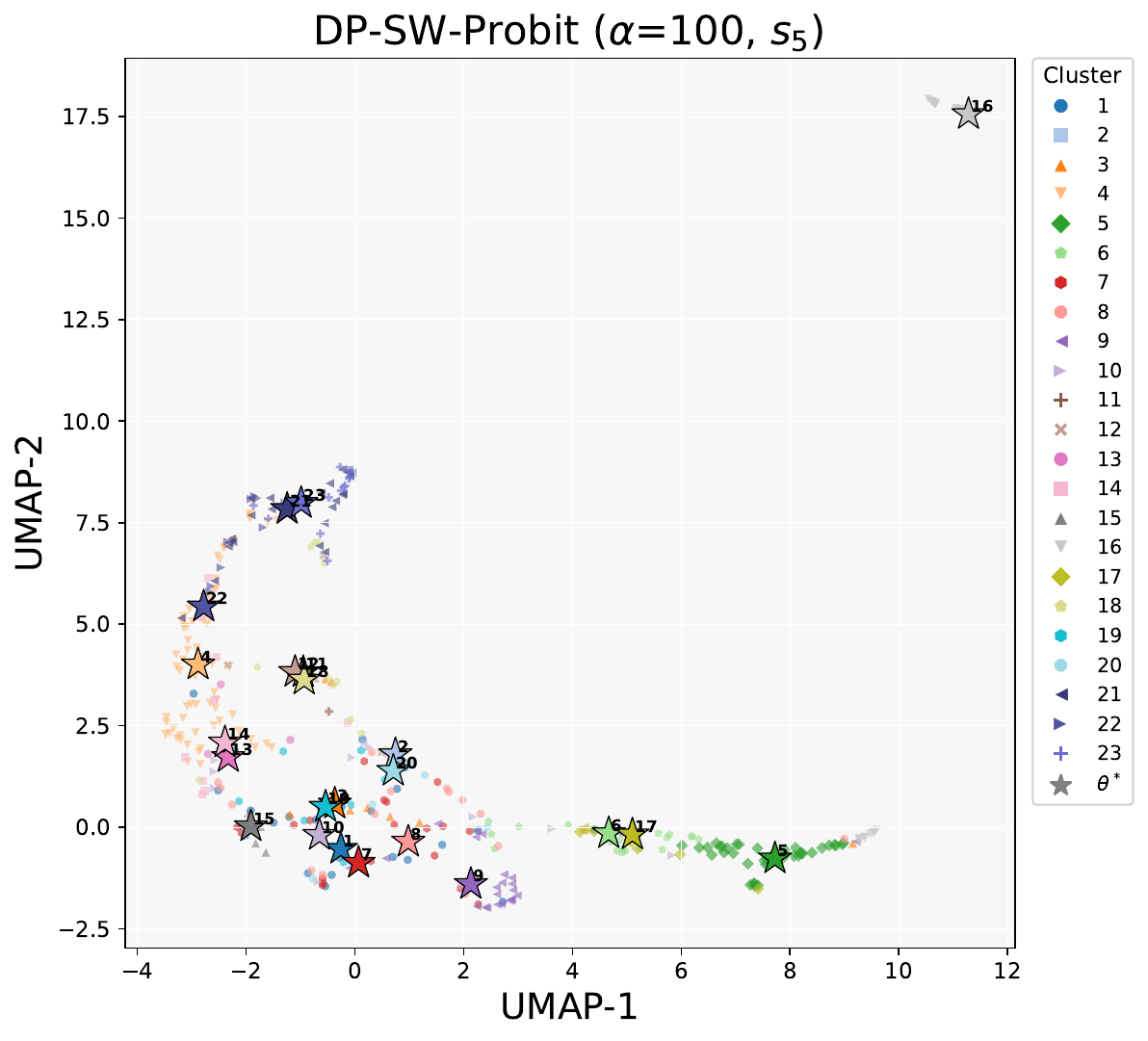} \\
    \end{tabular}
    \vspace{1em}
    \caption{UMAP embedding of patient trajectories based on pairwise $SW_2^2$ distances, for scenario $s_5$. Panels show, from top-left, the point-estimate partition $\widehat{S}$ under the DP-Probit and the  DP-SW-Probit  models with $\alpha=1$, $\alpha=10$, and $\alpha=100$. Points are colored by cluster membership; stars mark the estimated typical trajectories $\hat{\thb}_k^*$ (DP-SW-Probit).}
    \label{fig:Umap1}
\end{figure}

\begin{figure}[!tbp]
    \centering
    \begin{tabular}{c}
         \includegraphics[width=.8\linewidth]{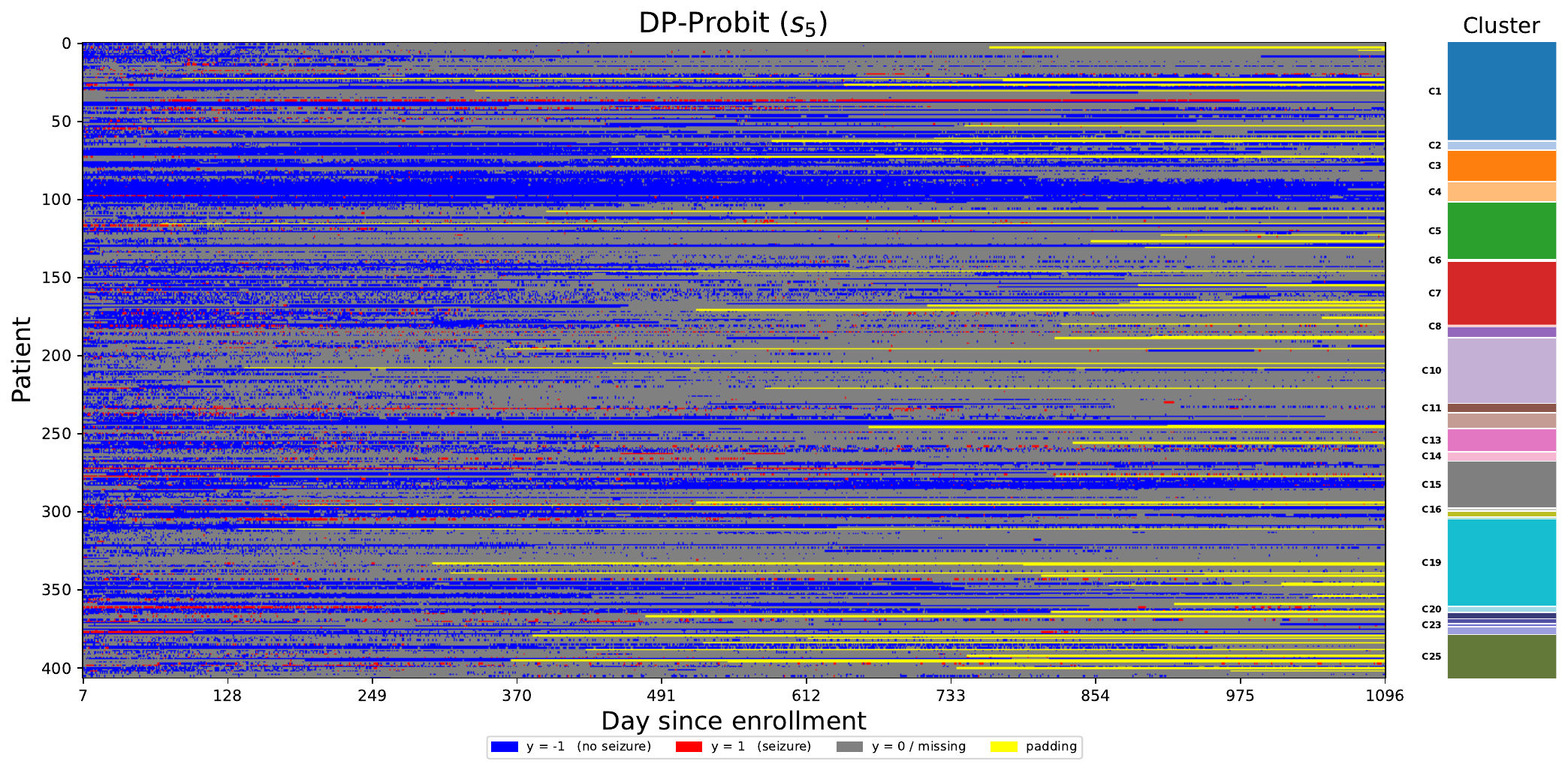}\\
         \includegraphics[width=.8\linewidth]{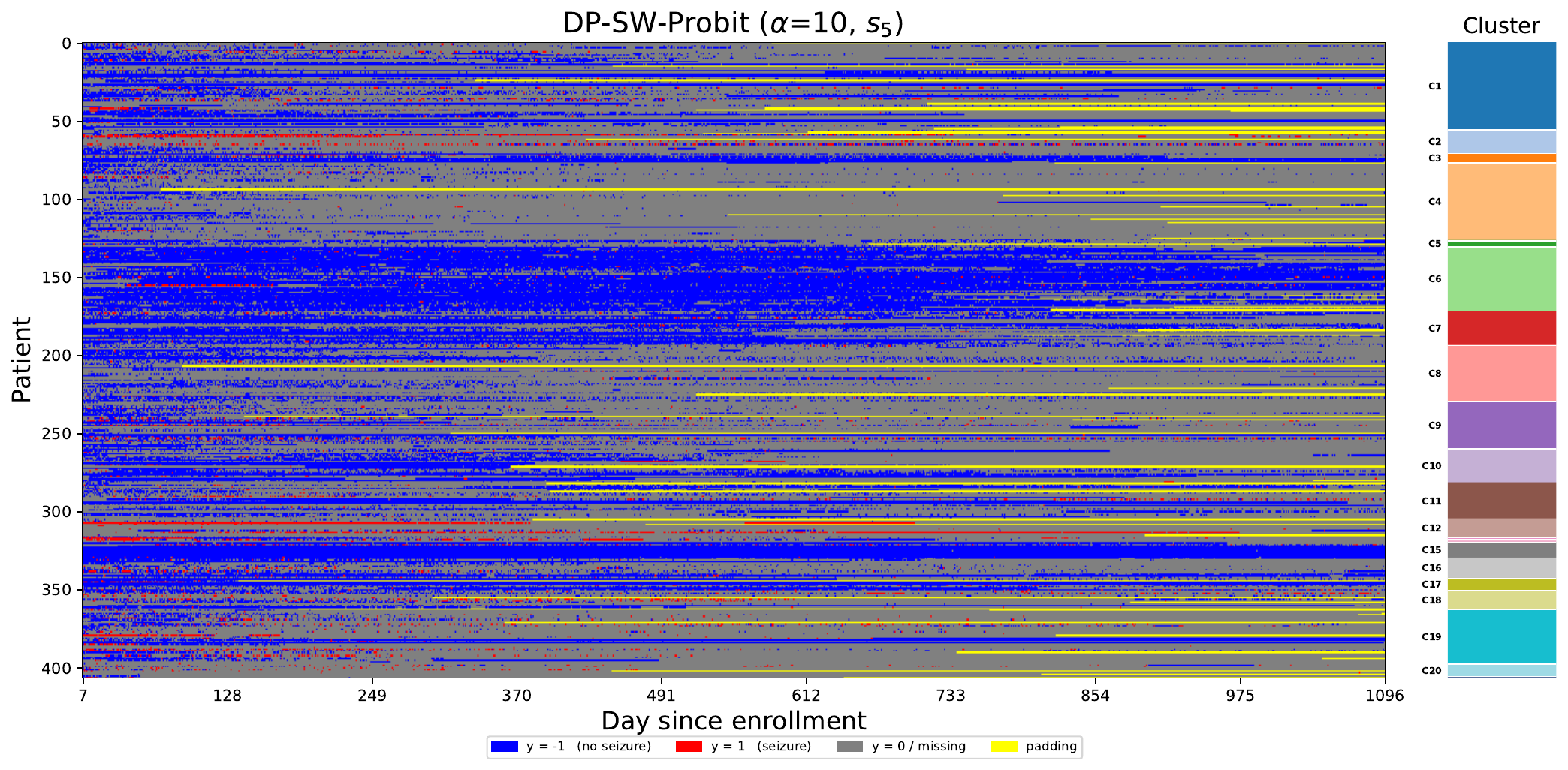} \\
         \includegraphics[width=.8\linewidth]{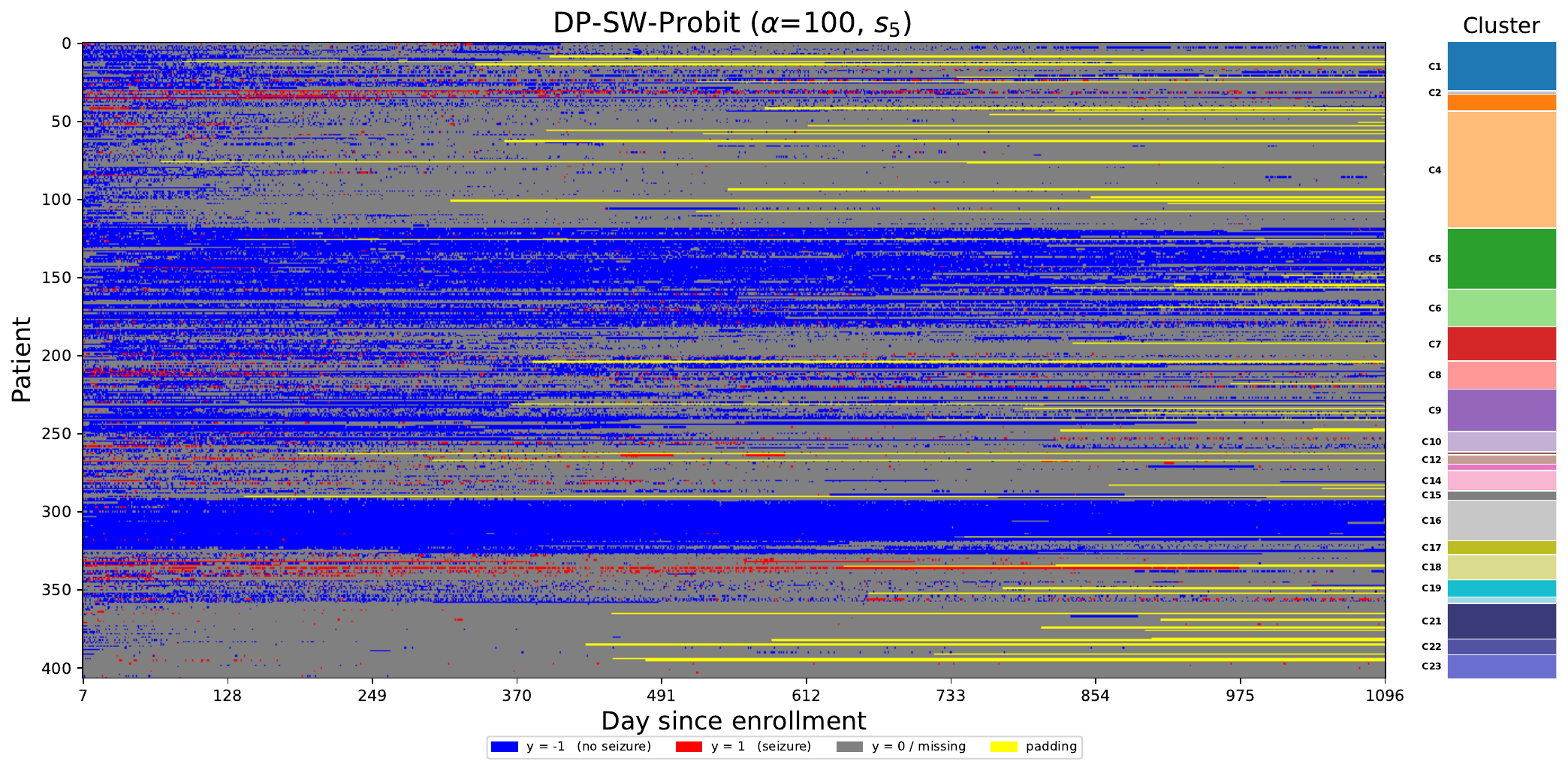} 
    \end{tabular}
    \vspace{1em}
    \caption{Heatmaps of full patient trajectories (rows), grouped by the point-estimate partition $\widehat{S}$ (scenario $s_5$), for  the DP-Probit (top) and  the DP-SW-Probit model with $\alpha=10$ (middle) and $\alpha=100$ (bottom).}
    \label{fig:Heat1}
\end{figure}

\begin{figure}[!tbp]
    \centering
    \begin{tabular}{c}
    \includegraphics[width=.8\linewidth]{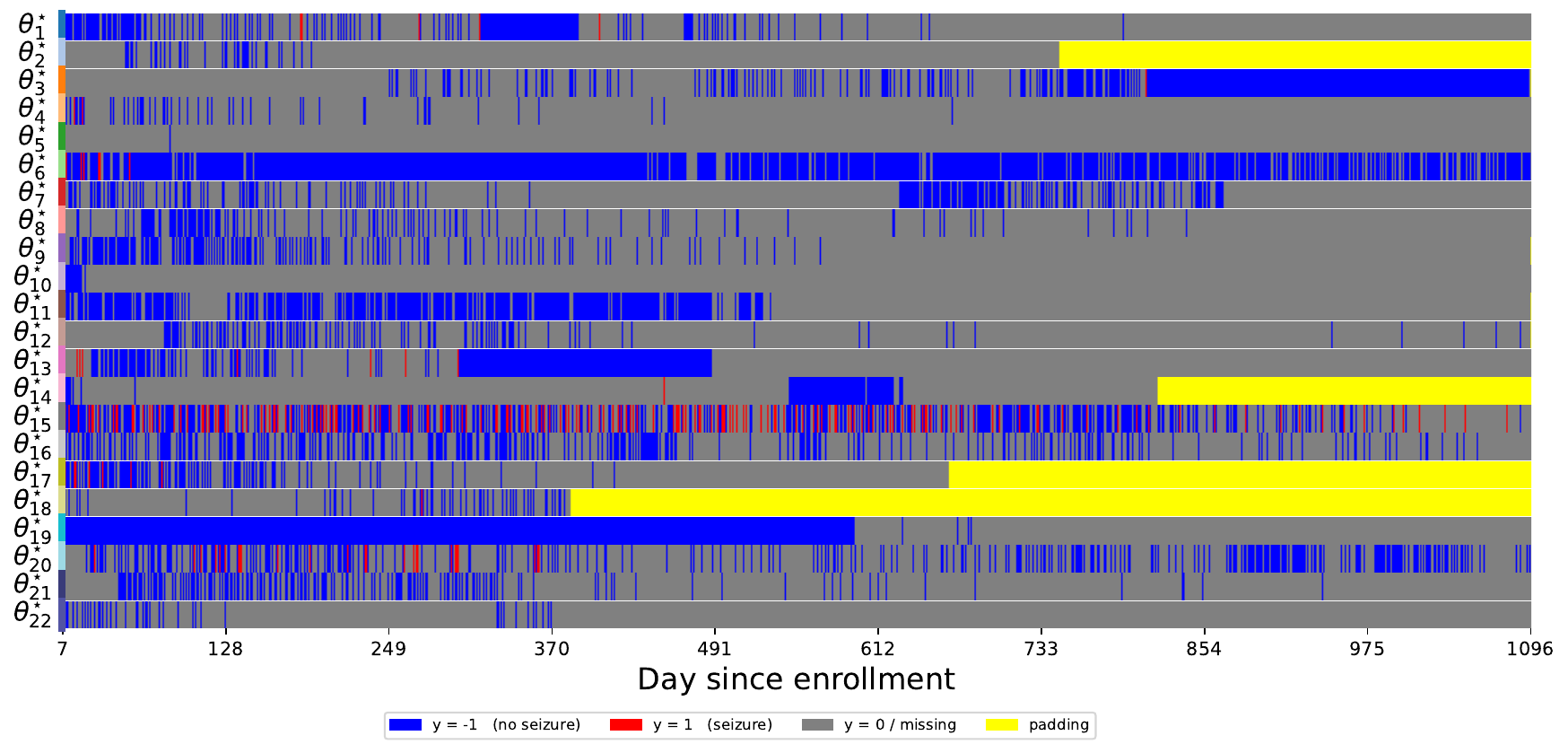}\\
         \includegraphics[width=.8\linewidth]{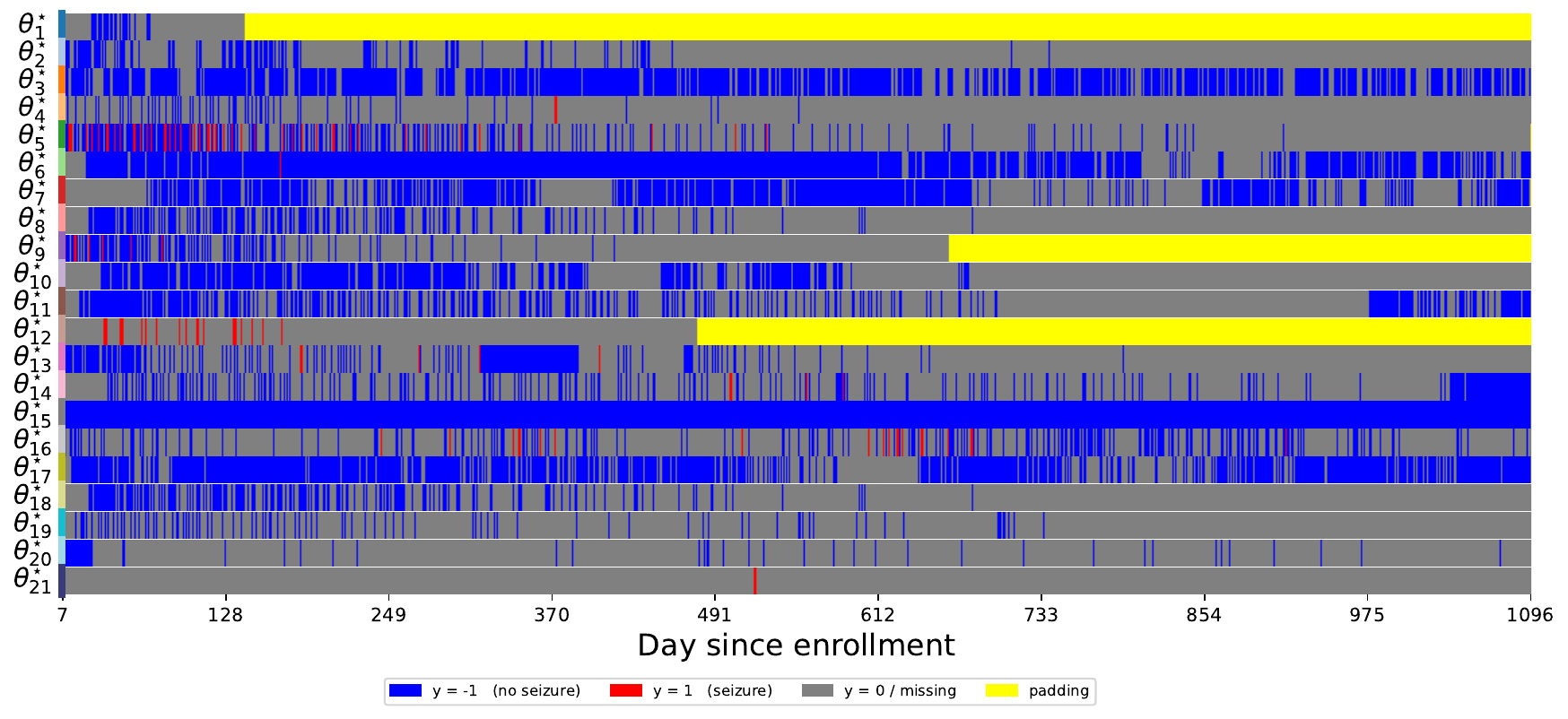}\\
         \includegraphics[width=.8\linewidth]{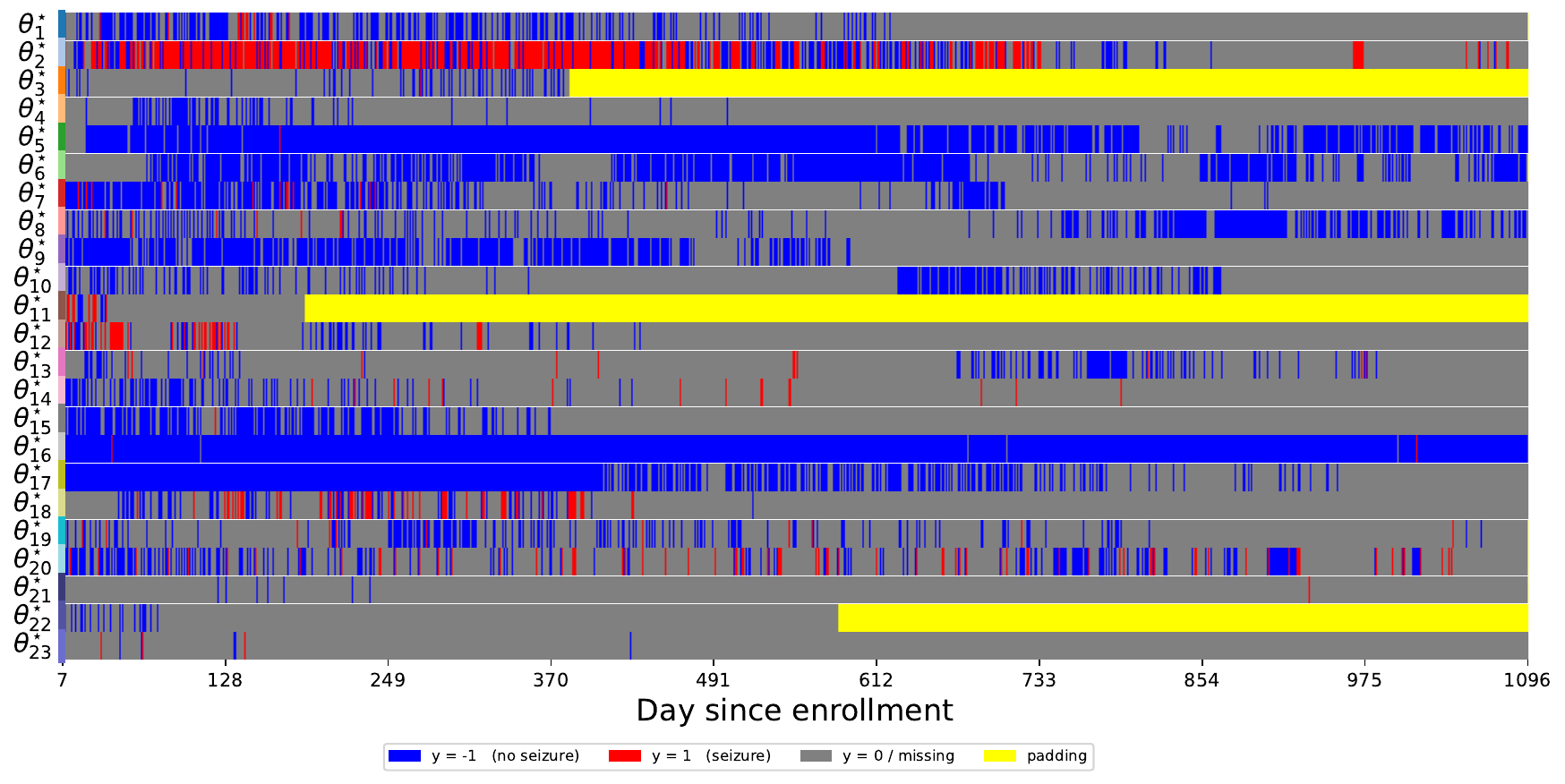}
    \end{tabular}
    \vspace{1em}
    \caption{Estimated typical-patient trajectories $\thb_k^*$ for each cluster under the DP-SW-Probit model (scenario $s_5$), for $\alpha=1$ (top), $\alpha=10$ (middle), and $\alpha=100$ (bottom).}
    \label{fig:Typical1}
\end{figure}

Figure~\ref{fig:Umap1} shows patient trajectories plotted as 
UMAP scores \citep{mcinnes2018umap} (using the precomputed pairwise
$SW_2^2$ matrix).
The figure also
shows reported point estimates of the random partition 
under the DP-Probit and under the DP-SW-Probit ($\alpha \in \{1,10,100\}$)
model, based on treatment covariates $s_5$.
For inference under the DP-SW-Probit model, we additionally
indicate (as stars) the typical
trajectories $\{\hat{\thb}^*\}_{1:K}$ associated with the reported random
partition. To plot $\hat{\thb}^*_{k}$ as  a typical trajectory, we recall the definition of $P_0^{(1)}$ in~\eqref{eq:base_measure} and plot the corresponding $\yb_i$. Compared to inference under the DP-Probit,
clusters under the proposed DP-SW-Probit model appear more homogeneous.
The typical trajectories provide a good representation of the patient
population under the SW geometry.
Figure~\ref{fig:Heat1} shows the actual data, as trajectories for all
patients grouped according to $\Shatb$.
Note again that patients within each group under the
DP-SW-Probit model appear more homogeneous compared to similar
trajectories under the DP-Probit model. 
Finally, Figure~\ref{fig:Typical1} shows the typical trajectories
under the DP-SW-Probit model which we recognize as good visual
summaries of patient trajectories in the cluster, as desired.

{\em Prediction.}
We now examine regression performance. We conduct in-sample prediction
using autoregressive prediction as described in
Section~\ref{subsec:prediction_and_clstering} with $\tau \in
\{1,3,7\}$ time steps ahead,
keeping the
time-dependent covariates as observed in the data. We evaluate the
Area Under the Receiver Operating Characteristic Curve (AUC)
for the prediction of the observed binary outcomes $y_{it}$.
Table~\ref{tab:insample_auc_full} reports AUC
scores for DP-Probit and for the proposed DP-SW-Probit. Overall, the proposed
model achieves higher AUC than DP-Probit across all
scenarios $s_1$--$s_5$ (using an increasing number of binary treatment
covariates) and across all choices of $w$. As expected, increasing $\tau$
leads to lower AUC scores with wider credible intervals. Larger values
of $\alpha$ slightly reduce AUC scores for the proposed model, as
they upweight the trajectory similarity loss. Adding more covariates
(from $s_1$ to $s_5$) does not generally improve AUC. However, the AUC
scores are generally high to guarantee the quality of prediction in
all scenarios.

\begin{table}[ht]
\centering
\footnotesize
\caption{In-sample AUC (posterior median, with 95\% credible interval)
  for one-, three-, and seven-day-ahead ($\tau$) prediction, comparing
  inference under DP-Probit (DP-P) vs. the  DP-SW-Probit (DP-SW-P) model across $\alpha
  \in \{1,10,100,1000\}$ and treatment-covariate scenarios
  $s_1$--$s_5$.} 
\label{tab:insample_auc_full}
\scalebox{0.9}{
  \hskip -.5cm
\begin{tabular}{lllccccc}
\toprule
$\tau$ & Method & $\alpha$ & s1 & s2 & s3 & s4 & s5 \\
\midrule
\multirow{7}{*}{1}
 & DP P    &               & 0.956 [0.936, 0.964] & 0.949 [0.926, 0.961] & 0.960 [0.949, 0.964] & 0.956 [0.941, 0.960] & 0.947 [0.935, 0.957] \\
 & DP-SW-P & 1     & 0.968 [0.964, 0.969] & 0.970 [0.966, 0.971] & 0.967 [0.964, 0.968] & 0.971 [0.967, 0.972] & 0.967 [0.958, 0.970] \\
 & DP-SW-P & 10    & 0.968 [0.964, 0.969] & 0.970 [0.966, 0.970] & 0.969 [0.964, 0.970] & 0.971 [0.968, 0.972] & 0.968 [0.956, 0.970] \\
 & DP-SW-P & 100   & 0.966 [0.963, 0.966] & 0.967 [0.964, 0.968] & 0.966 [0.963, 0.967] & 0.968 [0.965, 0.970] & 0.966 [0.956, 0.969] \\
 & DP-SW-P & 1000  & 0.964 [0.963, 0.964] & 0.965 [0.963, 0.965] & 0.966 [0.963, 0.967] & 0.969 [0.965, 0.970] & 0.965 [0.955, 0.968] \\
\addlinespace
\multirow{7}{*}{3}
 & DP P    &               & 0.944 [0.921, 0.957] & 0.942 [0.917, 0.953] & 0.950 [0.933, 0.956] & 0.940 [0.925, 0.944] & 0.929 [0.918, 0.945] \\
 & DP-SW-P & 1     & 0.960 [0.958, 0.963] & 0.963 [0.959, 0.965] & 0.961 [0.958, 0.962] & 0.965 [0.961, 0.966] & 0.960 [0.952, 0.963] \\
 & DP-SW-P & 10    & 0.960 [0.959, 0.962] & 0.961 [0.956, 0.963] & 0.963 [0.958, 0.965] & 0.964 [0.960, 0.966] & 0.962 [0.951, 0.964] \\
 & DP-SW-P & 100   & 0.956 [0.951, 0.959] & 0.960 [0.957, 0.962] & 0.958 [0.955, 0.960] & 0.961 [0.958, 0.964] & 0.960 [0.952, 0.963] \\
 & DP-SW-P & 1000  & 0.955 [0.950, 0.957] & 0.955 [0.946, 0.958] & 0.958 [0.952, 0.960] & 0.957 [0.942, 0.963] & 0.955 [0.946, 0.960] \\
\addlinespace
\multirow{7}{*}{7}
 & DP P    &               & 0.914 [0.887, 0.930] & 0.917 [0.889, 0.936] & 0.913 [0.896, 0.924] & 0.915 [0.900, 0.922] & 0.896 [0.883, 0.923] \\
 & DP-SW-P & 1     & 0.938 [0.930, 0.941] & 0.943 [0.938, 0.948] & 0.940 [0.935, 0.944] & 0.946 [0.940, 0.950] & 0.938 [0.920, 0.945] \\
 & DP-SW-P & 10    & 0.942 [0.938, 0.947] & 0.938 [0.930, 0.945] & 0.951 [0.944, 0.954] & 0.945 [0.939, 0.950] & 0.944 [0.925, 0.947] \\
 & DP-SW-P & 100   & 0.936 [0.925, 0.943] & 0.947 [0.940, 0.951] & 0.943 [0.938, 0.947] & 0.942 [0.936, 0.947] & 0.944 [0.934, 0.949] \\
 & DP-SW-P & 1000  & 0.939 [0.931, 0.943] & 0.934 [0.890, 0.943] & 0.941 [0.934, 0.947] & 0.926 [0.894, 0.946] & 0.937 [0.923, 0.946] \\
\bottomrule
\end{tabular}
}
\end{table}

{\em Treatment recommendation.}
Using treatment coding $s_5$, and the model with $\alpha=100$, we find
recommended treatment alternatives  based on patients' current
regimens. For each posterior sample, we search across all combinations
of the 5 binary treatment indicators and report the one that maximizes
the probability $1-\phat_{it}$, as defined in \eqref{phat}, of no
seizure at the next time step. 
Finally, we report the posterior probability of treatment
recommendation as the posterior probability for each treatment
being the maximizer.
Detail results appear in
Supplementary Section \ref{supp:trtReco}.
In summary the model reports benefits from switching to combination
therapy for almost all patients, across all clusters.
Inference is consistently suggesting to switch from mono- or
under-therapy toward regimens with at least two anti-seizure
medications.
Similarly, considering individual drugs (Levetiracetam, Lamotrigine,
other SCB, other ASM), inference reports benefits for switching to
more than one drug group for the same patient and visit, rather than
swapping one drug class for another. The gains seem to come from combining
complementary mechanisms, not substitution. For some clusters the
recommendation steers mainly toward Levetiracetam plus other SCB,
others toward Lamotrigine plus other ASM, suggesting the model tailors
the recommended polytherapy to each cluster's estimated response
profile rather than defaulting to one regimen for everyone. And within
a given patient's visit sequence, the recommended treatment
alternatives stays largely stable rather than flipping back and forth
from visit to visit, indicating a persistent target regimen rather
than posterior noise.

We also evaluate posterior estimates for the change in log probability
$(1-\phat_{it})$ of having no seizure at the next visit under the
recommended treatment, including 95\% posterior credible intervals for
$(1-\phat_{it})$. Details are reported in
Supplementary Section \ref{supp:trtReco}.
In summary, 
for most patient-visits we find a
small but consistently positive benefit from switching (posterior mean
in the $10^{-2}$--$10^{-1}$ range). A smaller group of patients shows
a much larger benefit, approaching or exceeding $10^0$--$10^1$. The
lower (2.5\%) credible bound sits close to zero for most
patient-visits, while the upper (97.5\%) bound pulls noticeably
further towards positive gains.
Aggregating within clusters reveals substantial
heterogeneity across clusters. The largest cluster, C4, together with
C5, C9, and C16, shows an average benefit close to zero throughout. C7
and C14 show a modest, sustained positive benefit (posterior mean
roughly $10^{-1}$), with C14 remaining credibly positive over part of
its range. C1, C6, and C8 show a similar modest positive mean only
over part of their observed range, fading toward zero thereafter.

\subsection{Prediction of Future Outcomes}
\label{subsec:prediction}
% \subsubsection{Setting up the comparisons}
We evaluate predictive performance under two complementary hold-out
schemes: a \textit{temporal} hold-out, in which every patient
contributes both training and test observations, and a
\textit{patient} hold-out, in which a subset of patients is withheld
entirely from model fitting. Both schemes rely on the same underlying
rule for partitioning a single patient's trajectory into a training
portion and a test portion, which we describe first. The many details
of setting up the comparison reflect the complex nature of the data.
Details are  reported in
Supplementary Section \ref{supp:prediction}, including in
particular details of the setup for the comparison.
In summary, inference under the proposed DP-SW-Probit model reports
higher AUC than the same inference under the DP-Probit, under temporal
as well as under patient hold-out, across all scenarios of coding
treatment indicators ($s_1$ through $s_5$), and across time horizons
$\tau=1$ through $\tau=7$. For out-of-sample prediction, under
temporal hold-out depending on
time horizon and coveriate coding, under DP-Probit AUC varies between
0.826 and 0.920. By comparison the same under the DP-SW-Probit varies
between 0.870 and 0.934.
Under patient hold-out we find under the DP-Probit AUC values between
0.563 and 0.850, while the same under the DP-SW-Probit are between
0.745 and 0.892.

\section{Conclusion}
\label{s:discuss}

We developed a generalized Bayesian framework for clustering and
regression analysis
with unaligned and irregularly observed longitudinal binary data.
The approach includes a two-component loss function.
The first component is a trajectory similarity
loss that uses the sliced Wasserstein distance between empirical
distributions of locally observed subsequences (“reads”), while the
other one is a regression loss based on an autoregressive probit model.
Besides accommodating the lack of alignment, the use of local reads
also allows inference for patients with varying duration of follow-up,
observation frequencies and missingness patterns.
A generalized Bayesian likelihood function integrates
the two components of the loss function without specifying a joint
generative model for outcomes, treatments, and missingness. 

In the data analysis, the proposed DP-SW-Probit model estimated fewer and
larger, more homogeneous clusters  when compared with the DP-Probit
model based on only the regression loss, and it showed better
predictive AUC in all considered hold-out scenarios.
% This effect was strongest for the patients
% completely left out of the model training: without any data to be used
% for fitting a regression likelihood on, these patients require
% clustering based on trajectory similarity only, and the results in
% Section~\ref{subsec:prediction} demonstrate how it leads to improved
% performance.
Also, the typical-patient trajectories $\thb_k^*$
were useful beyond clustering and prediction: since $P_0^{(1)}$ is supported on
actual trajectories, the $\thb_k^*$
correspond to actual patients, greatly facilitating clinical
interpretation.  An important feature in applications is that the typical trajectories implement inference for ``characteristic patterns", without the need to specify target patterns up front.

Several limitations remain. 
First, several fixed tuning parameters remain, including the 
read length which was fixed at $H=7$.
Second, missing values were treated as a third, neutral response
value, and were assumed to be missing at random.
However, it is known that epilepsy patients tend to
under-report seizure episodes.
Further limitations arise due to the nature of generalized
Bayes. Lacking a joint predictive distribution we implemented cluster
assignment for a new patient using the loss function only.
This appeared to work well in practice.
Finally, the model includes no inference on the relative weight
$\alpha$ in the two-component loss function.
However, we found results to be quite robust with respect to
$\alpha$, except for extreme values, and we have provided a
heuristic choice based on the scale of pairwise SW
distances between patients.

\appendix

\renewcommand{\thesection}{\Alph{section}}
\renewcommand{\thesubsection}{\Alph{section}.\arabic{subsection}}
\renewcommand{\thefigure}{S.\arabic{figure}}
\renewcommand{\thetable}{S.\arabic{table}}

\section{Results} 
\label{supp:results}
\subsection{Treatments Recommendation}
\label{supp:trtReco}
\begin{figure}[!tbp]
    \centering
    \begin{tabular}{c}
         \includegraphics[width=.91\linewidth]{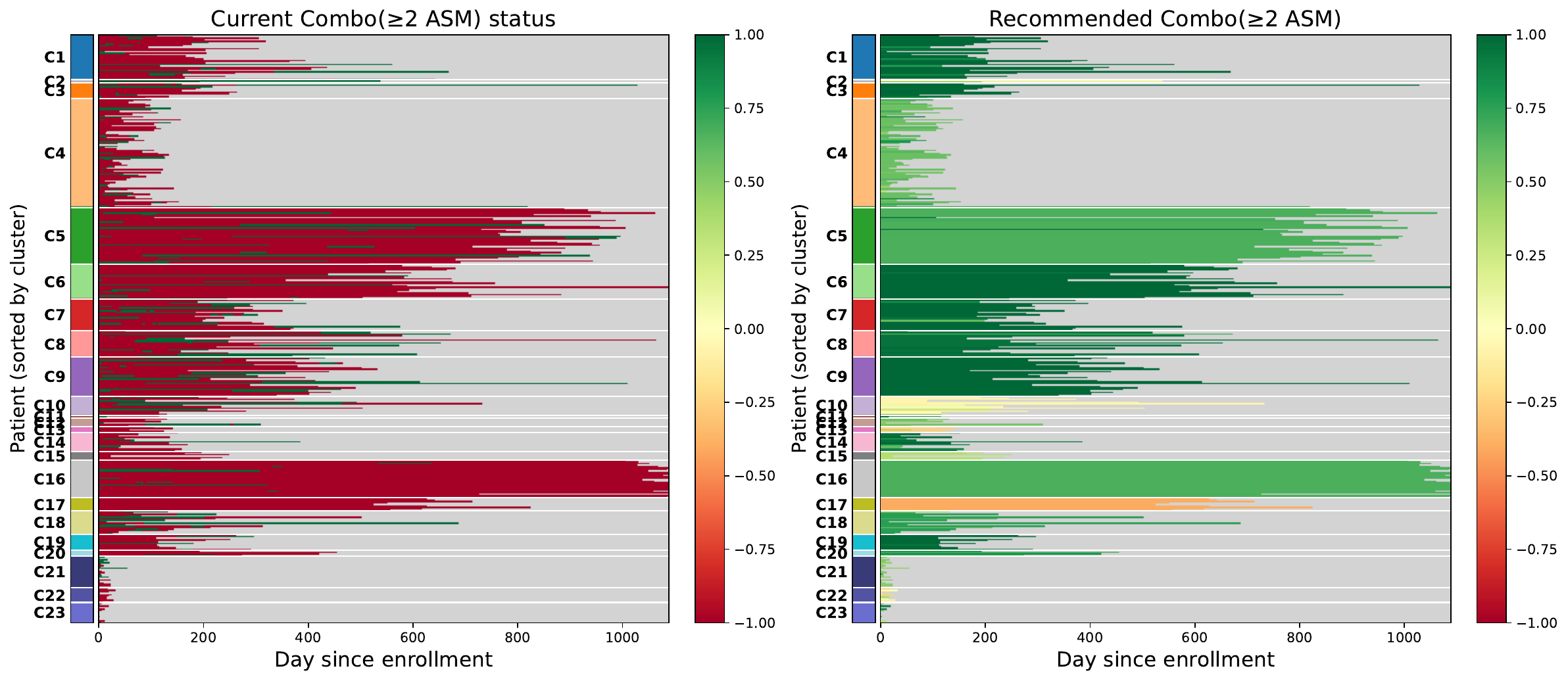} \\
         \includegraphics[width=.91\linewidth]{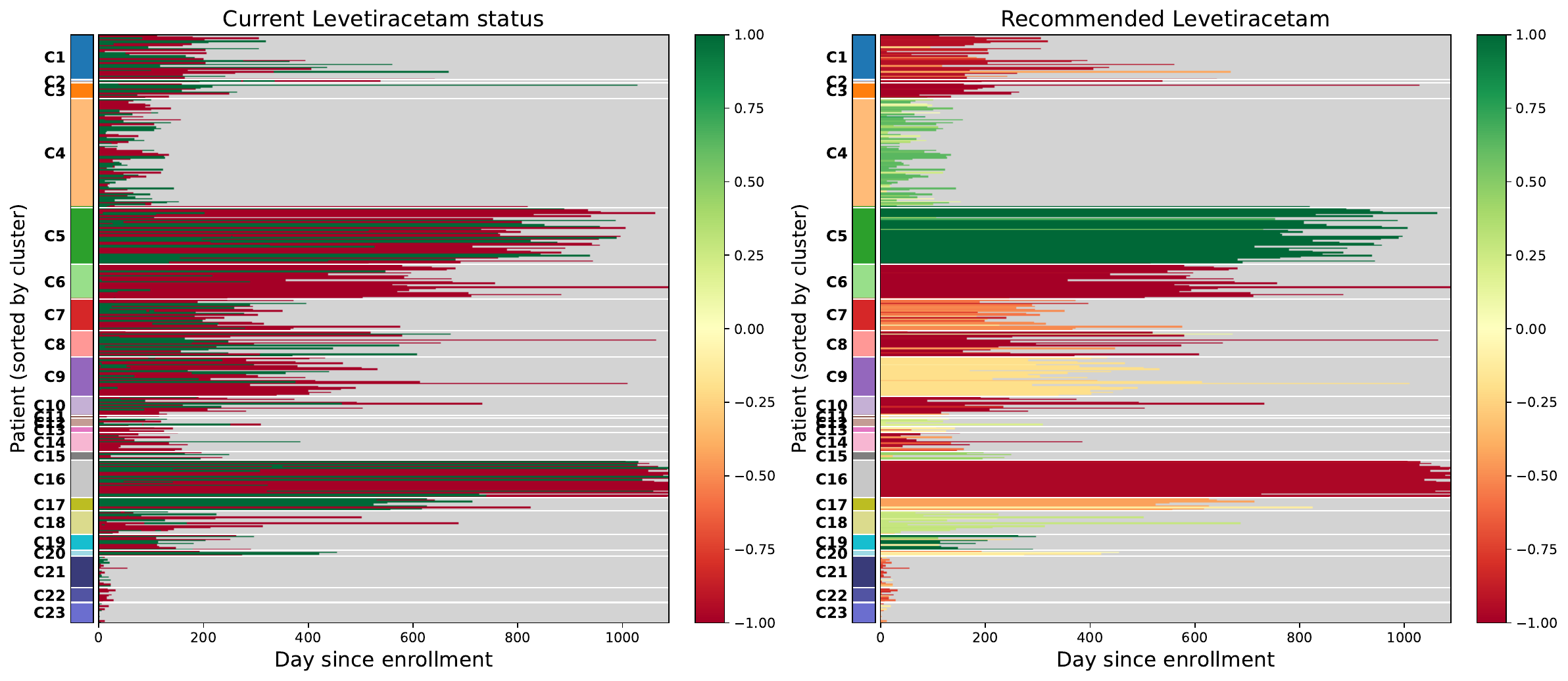} \\
          \includegraphics[width=.91\linewidth]{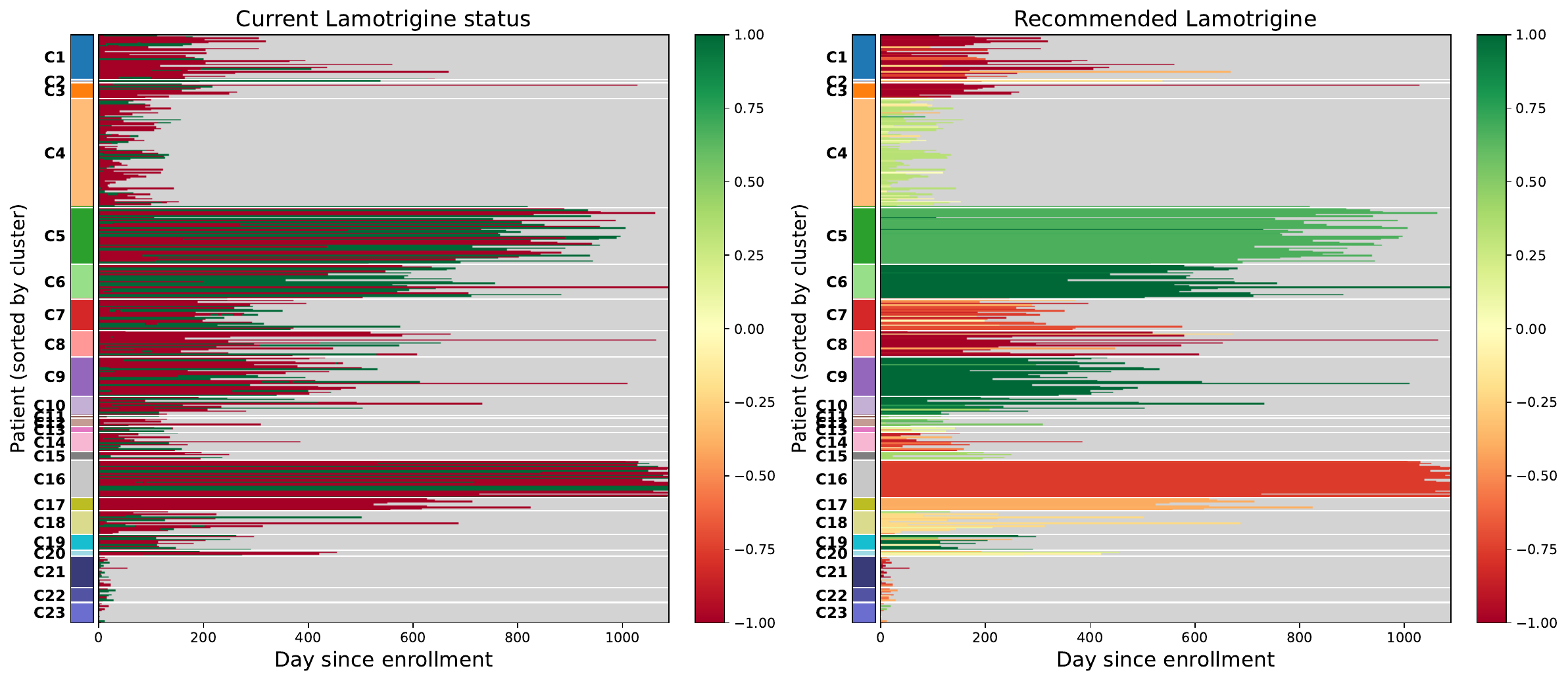}
    \end{tabular}
    \vspace{-1em}
    \caption{Observed versus model-recommended treatment assignment
      over time for combination therapy (top), Levetiracetam (middle),
      and Lamotrigine (bottom), for patients grouped by clusters under
      the DP-SW-Probit model ($\alpha=100$, scenario $s_5$). Under
      ``recommended'' (right column), green indicates the respective
      treatment is recommended. Under ``current treatment'' (left column)
      green/red means the treatment is assigned/not assigned.} 
    \label{fig:treament_1}
\end{figure}

\begin{figure}[!t]
    \centering
    \begin{tabular}{c}
         \includegraphics[width=1\linewidth]{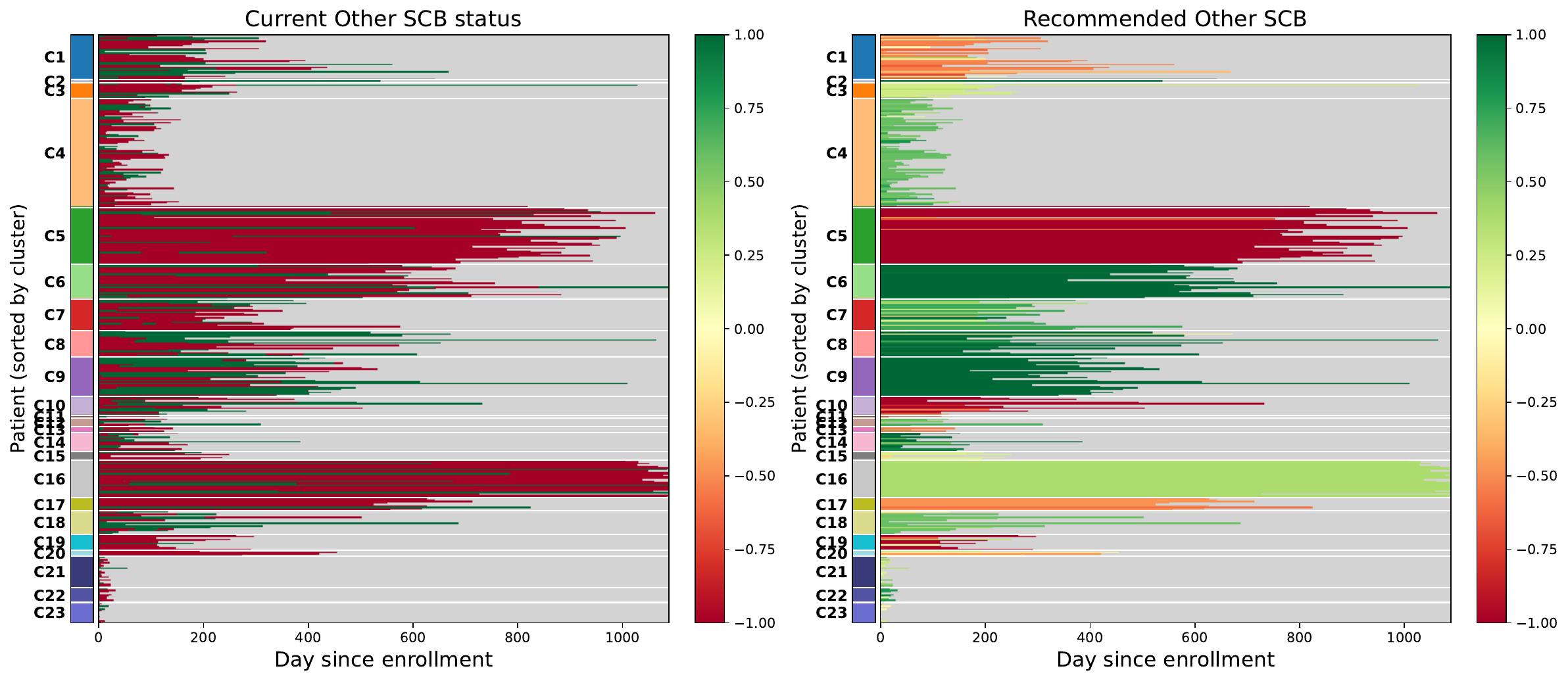} \\
         \includegraphics[width=1\linewidth]{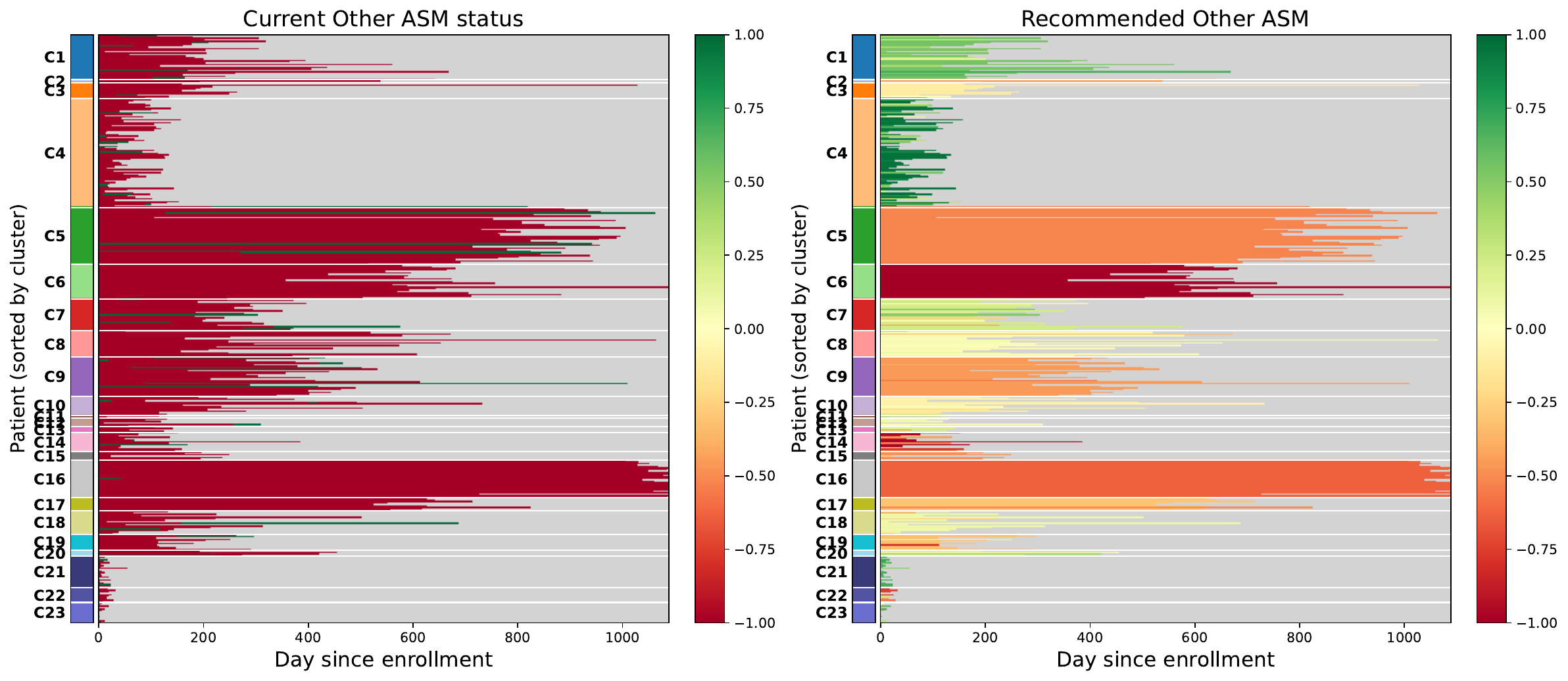}
    \end{tabular}
    \vspace{-1em}
    \caption{Same as Figure~\ref{fig:treament_1} for other sodium-channel blockers (top) and other anti-seizure medications (bottom).}
    \label{fig:treament_2}
\end{figure}

We report more details for the results on treatment recommendations
that were summarized in Section \ref{subsec:experiment_clustering}.
Using treatment coding $s_5$, and the model with $\alpha=100$, we find
recommended treatment alternatives  based on patients' current
regimens. For each posterior sample, we search across all combinations
of the 5 binary treatment indicators and report the one that maximizes
the probability $1-\phat_{it}$ of no seizure at the next time step.
% Finally, we report the posterior probability of treatment
% recommendation as the posterior probability for each treatment
% being the maximizer.
Figures~\ref{fig:treament_1}-\ref{fig:treament_2} show the current
treatments alongside the recommended treatment alternatives, with
patients grouped by $\widehat{S}$.
Almost every group stands to benefit from combination therapy.  

The combo panel is noticeably greener (recommended) than the observed
panel across nearly all clusters, which means that the model is
consistently pushing patients from mono- or under-therapy toward
regimens with at least two anti-seizure medications. The individual
drug-level panels (Levetiracetam, Lamotrigine, other SCB, other ASM)
tell a similar story. The recommendation often turns green for more
than one drug group for the same patient and visit, rather than
swapping one drug class for another. The gains seem to come from
combining complementary mechanisms, not substitution. For some
clusters the recommendation steers mainly toward Levetiracetam plus
other SCB, others toward Lamotrigine plus other ASM, suggesting the
model tailors the recommended polytherapy to each cluster's estimated
response profile rather than defaulting to one regimen for
everyone. And within a given patient's visit sequence, the recommended
treatment alternatives stays largely stable rather than flipping back
and forth from visit to visit, indicating a persistent target regimen
rather than posterior noise.

\begin{figure}[!t]
    \centering
    \begin{tabular}{c}
         \includegraphics[width=1\linewidth]{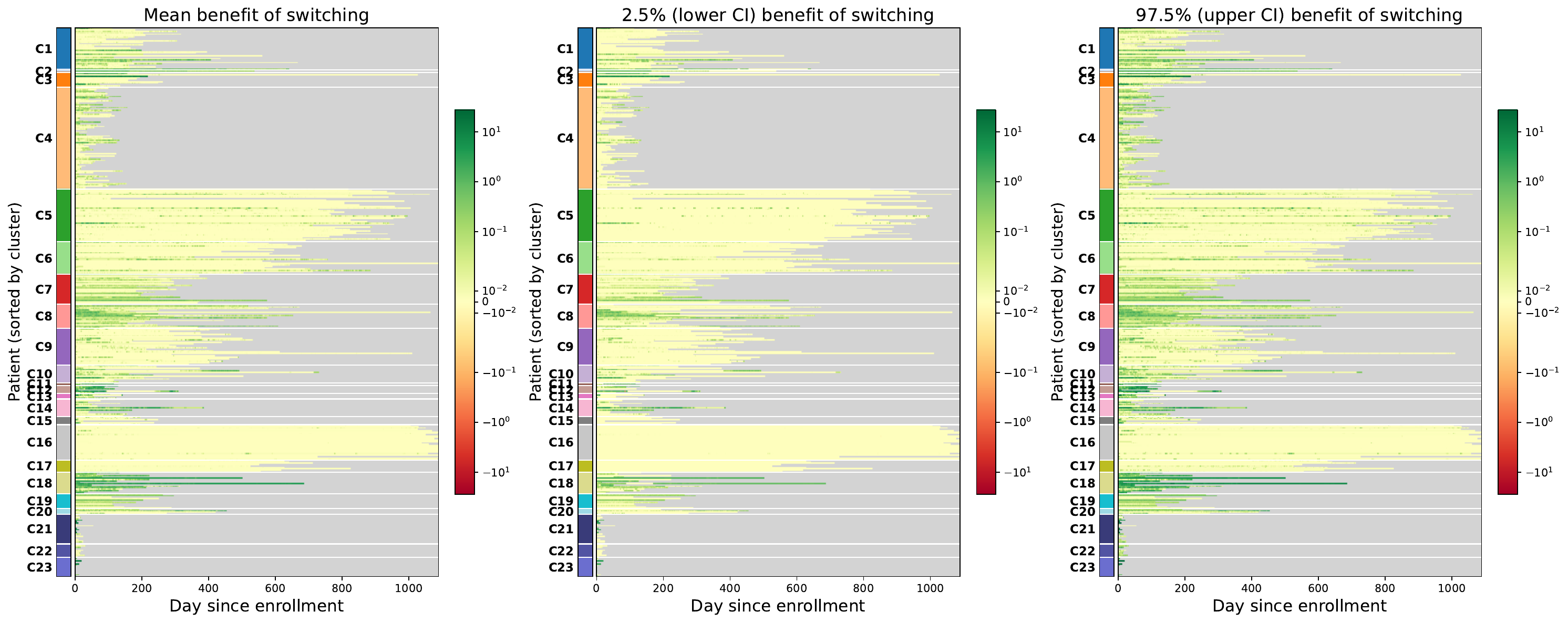}
         \\
          \includegraphics[width=1\linewidth]{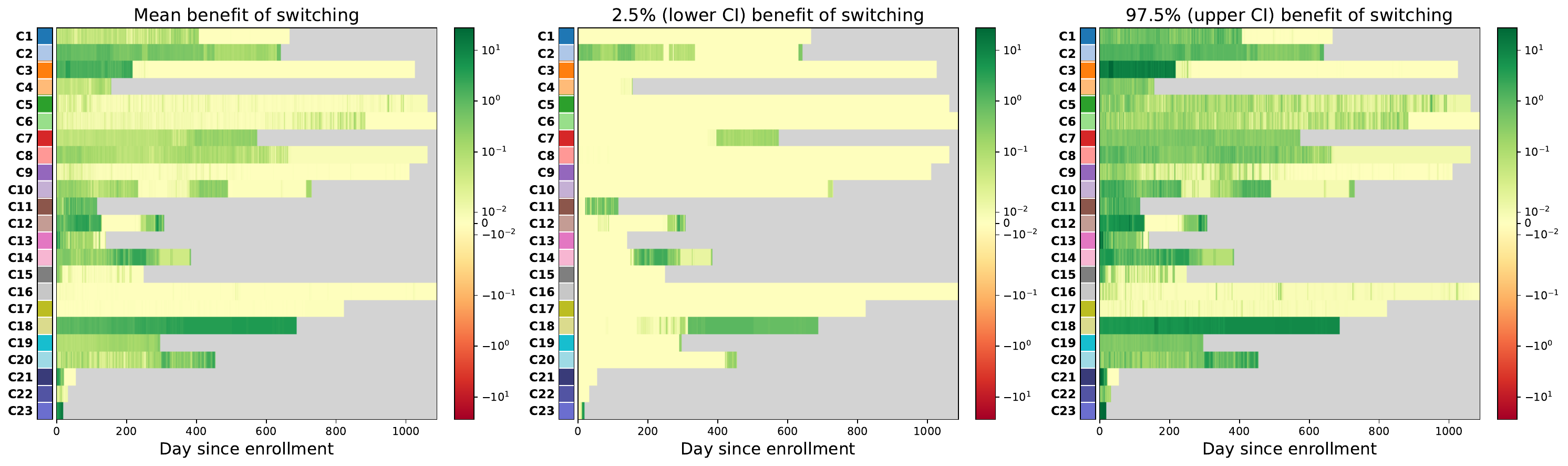}
    \end{tabular}
    \vspace{-1em}
    \caption{Posterior expectation (with 95\% credible interval) of the change in log-probability of no seizure at the next visit under the recommended versus observed treatment regimen, at the patient level (top) and averaged within cluster (bottom), under DP-SW-Probit ($\alpha=100$, scenario $s_5$).}
    \label{fig:benefit}
\end{figure}

Given the recommended treatment, we calculate the change in log
probability $(1-\phat_{it})$ of having no seizure at the next
visit. Figure~\ref{fig:benefit} reports the posterior expectation of
this summary along with the 95\% credible interval. The figure also
breaks out the average within each cluster. At the patient level
(first row in Figure~\ref{fig:benefit}), for most patient-visits we find a
small but consistently positive benefit from switching (posterior mean
in the $10^{-2}$--$10^{-1}$ range). A smaller group of patients shows
a much larger benefit, approaching or exceeding $10^0$--$10^1$. The
lower (2.5\%) credible bound sits close to zero for most
patient-visits, while the upper (97.5\%) bound pulls noticeably
further into positive territory. Aggregating within cluster (second
row in Figure~\ref{fig:benefit}, $\alpha = 100$) reveals substantial
heterogeneity across clusters. For cluster C18 we find a sustained, credibly
positive benefit across nearly its entire observed range, with the
lower credible bound remaining above zero. C3 shows a large positive
point estimate early in its range, but the credible interval includes
zero throughout, so this apparent early benefit is not statistically
distinguishable from no effect. The largest cluster, C4, together with
C5, C9, and C16, shows an average benefit close to zero throughout. C7
and C14 show a modest, sustained positive benefit (posterior mean
roughly $10^{-1}$), with C14 remaining credibly positive over part of
its range. C1, C6, and C8 show a similar modest positive mean only
over part of their observed range, fading toward zero thereafter.

\subsection{Prediction of Future Outcomes}
\label{supp:prediction}
We report more details for the evaluation of predictive inference that
was summarized in Section \ref{subsec:prediction}.

We evaluate predictive performance under two complementary hold-out
schemes: a \textit{temporal} hold-out, in which every patient
contributes both training and test observations, and a
\textit{patient} hold-out, in which a subset of patients is withheld
entirely from model fitting. Both schemes rely on the same underlying
rule for partitioning a single patient's trajectory into a training
portion and a test portion, which we describe first. The many details of setting up the comparison reflect the complex nature of the data.

\paragraph{Within-patient temporal split} We fix the number of
held-out tracked days $n_h$ (we use $n_h=10$). From
Section~\ref{section:model}, the regression design vector $\xb_{it}$
requires a complete lag window of length $H$, so evaluation is only
defined for $t \in \mathcal{A}_i := \{H+1,\ldots,T_i\}$. Let 
$$
    \OO_i^{H} \;=\; \OO_i \cap \mathcal{A}_i
    \;=\; \big\{ o_i^{(1)} < o_i^{(2)} < \cdots < o_i^{(|\OO_i^{H}|)} \big\}
$$
denote the tracked days of patient $i$ that also admit a full $H$-day lag window, sorted in increasing order. Patient $i$ is eligible for a temporal split if $|\OO_i^{H}| \ge n_h$ (otherwise patient $i$ is excluded from the corresponding evaluation). For an eligible patient, we define the split day as the $n_h$-th tracked day before the end of follow-up 
$
    t_i^{\ast} \;=\; o_i^{\left(|\OO_i^{H}| - n_h + 1\right)}.
$
This choice guarantees that the terminal segment $\{t \in \mathcal{A}_i : t \ge t_i^{\ast}\}$ contains exactly $n_h$ tracked days while the initial segment $\{t \in \mathcal{A}_i : t < t_i^{\ast}\}$ contains everything observed earlier. We write
$$
    \mathcal{A}_i^{\mathrm{tr}} = \{t \in \mathcal{A}_i : t < t_i^{\ast}\},
    \qquad
    \mathcal{A}_i^{\mathrm{te}} = \{t \in \mathcal{A}_i : t \ge t_i^{\ast}\},
    \qquad
    |\OO_i \cap \mathcal{A}_i^{\mathrm{te}}| = n_h .
$$
The outcome sequence $\yb_i$, treatment sequence $\zb_i$, lagged
design vectors $\{\xb_{it}\}$, and read set $\mathcal{R}(\yb_i)$
(Section \ref{subsec:trajectory_similarity_loss}) are each restricted
to $\mathcal{A}_i^{\mathrm{tr}}$ when forming the training data, and
to $\mathcal{A}_i^{\mathrm{te}}$ when forming the held-out test
data. In particular, the reads used to compute $\DD(\yb_i,\thb_k)$
during posterior inference are built only from $y_{it}$, $t \in
\{1,\ldots,t_i^{\ast}-1\}$, so that no held-out observation enters
either the regression loss $L_2$ or the trajectory similarity loss
$L_1$. 

Next we define two different hold-out schemes (temporal vs.
patients) and different evaluation sets (in-sample vs.\
out-of-sample). Combinations of the two give rise to five different
sets of comparisons summarized in Table \ref{table:comparison} and
labeled 1(a) through 2B(b). For easier reference we refer to the
corresponding lines in the table as ``\#x"  in the discussion below.
\begin{table}[ht]
  \footnotesize
  \caption{Data splits (hold-out schemes) in the evaluation of predictive inference and their use for inference. The last column indicates the table that reports the AUC for prediction classification of the corresponding patient responses as $y=1$ or $y=-1$ (in in-sample fitting and out-of-sample prediction). }
  \begin{center}
  \begin{tabular}{ll|cccc}
    \hline
    & & \multicolumn{3}{c}{used for}  \\
    \cline{3-5}
      & & posterior & cluster & covariates \\
    \multicolumn{2}{l}{Data split}  & inference & assignment & in probit & Table\\
    \hline
    \multicolumn{5}{c}{
    1. Temporal hold-out ($n=381$) (within patient split)}\\[2pt]
    1(a): in-sample & $t<\ts_i$                  & x & x & x &\ref{tab:in_auc} \\
    1(b): out-of-sample & $t \ge \ts_i$          &   &   & x &\ref{tab:out_auc}\\[8pt]
    \multicolumn{5}{c}{
    2. Patient hold-out (splitting the cohort)}\\[2pt]
    \multicolumn{5}{l}{2A. in-sample $M^\tr=366$} \\
    ~~ (no temporal split) &            & x & x & x&\ref{tab:patients_in_auc}\\[4pt]
   
    \multicolumn{5}{l}{2B. out-of sample, $M^\te=39$}\\
     ~~~ 2B(a) history& $t< \ts_i$   &   & x & x\\
    ~~~ 2B(b) target, & $t \ge \ts_i$&   &   & x&\ref{tab:patients_out_auc}\\
    \hline
  \end{tabular}
  \end{center}
  \label{table:comparison}
\end{table}
\paragraph{Temporal hold-out (\#1)} Every patient
$i=1,\ldots,M$ that is eligible under the rule above is split
individually at $t_i^{\ast}$. In total we find 381 of 407 eligible patients
(26 excluded), $n_h=10$ tracked test days per patient, with
test-segment lengths 
spanning between 11 and 1088 days (median 319), and training-segment spans
2 to 1079 days. The
model is fit using only $\{(\yb_i^{\mathrm{tr}}, \zb_i^{\mathrm{tr}},
\bb_i)\}_{i=1}^{M}$, where $\yb_i^{\mathrm{tr}} = (y_{it})_{t \in
  \mathcal{A}_i^{\mathrm{tr}}}$ and similarly for
$\zb_i^{\mathrm{tr}}$. The \underline{\textit{In-sample}} AUC (\#1(a))(Table~\ref{tab:in_auc})
is computed by comparing the fitted probabilities
$\phat_{it} = \Phi(\betab_{S_i}^{*\top}\xb_{it})$ from \eqref{phat} against $y_{it}$, $t \in \OO_i
\cap \mathcal{A}_i^{\mathrm{tr}}$. The \underline{ \textit{Out-of-sample}} AUC (\#1(b))
(Table~\ref{tab:out_auc}) is computed by running the autoregressive
forecast of Section~ \ref{subsec:prediction_and_clstering} forward from
$t_i^{\ast}-1$ and comparing $\phat_{i,t}$ against $y_{it}$ for $t
\in \OO_i \cap \mathcal{A}_i^{\mathrm{te}}$, using the treatment
history $\zb_i$ (assumed known over the forecast horizon) and the
cluster assignment $S_i$ obtained from fitting on
$\mathcal{A}_i^{\mathrm{tr}}$ alone.

\paragraph{Patient hold-out (\#2)} We first partition the cohort itself. We draw a uniform random permutation of $\{1,\ldots,M\}$ and set aside a fraction $f_{\mathrm{te}}$ (we use $f_{\mathrm{te}}=0.10$) as \textit{test-candidate} patients, with the remainder forming the \textit{training} patients:
$$
    M_{\mathrm{cand}} = \max(1,\ \mathrm{round}(f_{\mathrm{te}} M)), \qquad
    \mathcal{M}^{\mathrm{tr}} \cup\, \mathcal{M}^{\mathrm{cand}} = \{1,\ldots,M\}, \qquad
    |\mathcal{M}^{\mathrm{cand}}| = M_{\mathrm{cand}} .
$$
{\it Training patients} (\#2A)  with $i \in \mathcal{M}^{\mathrm{tr}}$ retain their
entire trajectory $(\yb_i, \zb_i, \bb_i)$ and are used in full to fit
the model as in Section~ \ref{subsec:posterior_inference}. Each test
candidate $i \in \mathcal{M}^{\mathrm{cand}}$ is then subjected to the
within-patient temporal split above: candidates with $|\OO_i^{H}| <
n_h$ are dropped, leaving the final test set
$\mathcal{M}^{\mathrm{te}} \subseteq \mathcal{M}^{\mathrm{cand}}$
(with $|\mathcal{M}^{\mathrm{te}}| = M_{\mathrm{te}} \le
M_{\mathrm{cand}}$). For $i \in \mathcal{M}^{\mathrm{te}}$ (\#2B), the
pre-split segment $\mathcal{A}_i^{\mathrm{tr}}$ (\textit{history}, \#2B(a)) is
never used in fitting. It is instead used only at prediction time to
assign patient $i$ to a cluster via the score $S_{i}=\arg\max_k n_k
\LL(\yb_i^{\mathrm{tr}};\bm\phi_k^{*})$ of
Section~ \ref{subsec:prediction_and_clstering}, treating $i$ as a new
patient $i^*$. The post-split segment $\mathcal{A}_i^{\mathrm{te}}$
(\textit{target}, exactly $n_h$ tracked days, \#2B(b)) is held out entirely for
evaluation and is never seen by the sampler or by the clustering
step. Overall, we obtain $ M_{\mathrm{cand}}=41$ candidates with 2
dropped which leads to $M_{\mathrm{te}}=39$ held-out patients,
$M_{\mathrm{tr}}=366$ training patients. History span 2–1079 days
(median 548) for held-out patients, target span 11–1088 days (median
435), and full training-patient trajectories span up to 1090 days
(median 1090).

\underline{\textit{In-sample}} AUC (\#2A)
(Table~\ref{tab:patients_in_auc}) is computed on training patients $i
\in \mathcal{M}^{\mathrm{tr}}$ exactly as in the temporal-hold-out in-sample case.
\underline{\textit{Out-of-sample}} AUC (\#2B(b))
(Table~\ref{tab:patients_out_auc}) is computed on the target segments
of the held-out test patients $i \in \mathcal{M}^{\mathrm{te}}$, so
that this quantity reflects predictive performance for patients who
contributed \emph{no} data whatsoever to model fitting, as opposed to
the temporal hold-out, where every patient contributes a training
portion. 

\subsubsection{Results}
Tables~\ref{tab:in_auc}--\ref{tab:patients_out_auc} report AUC under
all four combinations of hold-out scheme (temporal vs.\ patient) and
evaluation set (in-sample vs.\ out-of-sample).
In the next few paragraphs we summarize the results in these tables.

\paragraph{In-sample predictions}
Under the \underline{temporal hold-out}, in-sample AUC (\#1(a), Table~\ref{tab:in_auc}) is uniformly high for
both methods (DP-Probit in the low-to-mid 0.95s, DP-SW-Probit
typically 0.95--0.97), and the proposed model retains a small but
consistent edge over DP-Probit across scenarios $s_1$--$s_5$. Here the
effect of $\alpha$ is comparatively small and does not follow a
single consistent direction across scenarios, indicating that
in-sample fit is not very sensitive to how much weight is placed on
the trajectory similarity loss once a patient's own training history
is available. Under the \underline{patient hold-out}, in-sample AUC
(\#2A, Table~\ref{tab:patients_in_auc}, computed on training patients only)
shows a substantially larger gap in favor of DP-SW-Probit (typically
0.96--0.97 versus 0.90--0.95 for DP-Probit at $\tau=1$), and this gap
widens further at $\tau=7$. The performance across $\alpha$ is again
fairly stable except at the largest value $\alpha=1000$, which shows
both a drop in mean AUC and markedly wider credible intervals at
$\tau=7$ (e.g., $s_1$: 0.922 [0.850, 0.946]), suggesting that
overweighting the trajectory loss can destabilize the regression fit
on longer horizons.

% Reorganized: rows = tau (grouped) x Method, columns = Treatment (s1-s5)
\begin{table}[!t]
\centering
\footnotesize
\caption{Temporal hold-out, in-sample AUC (\#1(a) in Table~\ref{table:comparison}) (posterior mean, with 95\% credible interval) for inference under the DP-Probit and the  DP-SW-Probit models across $\alpha \in \{1,10,100,1000\}$, evaluated on the training portion $\mathcal{A}_i^{\mathrm{tr}}$ of each patient's trajectory at horizons $\tau \in \{1,3,7\}$.}
\label{tab:in_auc}
\scalebox{0.8}{
\begin{tabular}{llccccc}
\toprule
$\tau$ & Method & s1 & s2 & s3 & s4 & s5 \\
\midrule
\multirow{7}{*}{1}
 & DP Probit                    & 0.950 [0.939, 0.958] & 0.960 [0.943, 0.967] & 0.951 [0.938, 0.957] & 0.958 [0.943, 0.967] & 0.950 [0.939, 0.958] \\
 & DP-SW-Probit ($\alpha=1$)    & 0.957 [0.954, 0.960] & 0.967 [0.965, 0.968] & 0.965 [0.963, 0.966] & 0.967 [0.965, 0.968] & 0.951 [0.948, 0.955] \\
 & DP-SW-Probit ($\alpha=10$)   & 0.956 [0.953, 0.960] & 0.964 [0.962, 0.965] & 0.964 [0.962, 0.966] & 0.966 [0.965, 0.967] & 0.955 [0.953, 0.957] \\
 & DP-SW-Probit ($\alpha=100$)  & 0.958 [0.955, 0.960] & 0.964 [0.962, 0.964] & 0.965 [0.964, 0.966] & 0.961 [0.957, 0.964] & 0.951 [0.948, 0.953] \\
 & DP-SW-Probit ($\alpha=1000$) & 0.965 [0.965, 0.966] & 0.966 [0.965, 0.967] & 0.967 [0.965, 0.968] & 0.967 [0.965, 0.968] & 0.951 [0.948, 0.953] \\
\addlinespace
\multirow{7}{*}{3}
 & DP Probit                    & 0.939 [0.927, 0.948] & 0.947 [0.929, 0.959] & 0.941 [0.928, 0.947] & 0.949 [0.931, 0.959] & 0.938 [0.926, 0.947] \\
 & DP-SW-Probit ($\alpha=1$)    & 0.949 [0.944, 0.954] & 0.959 [0.958, 0.961] & 0.957 [0.954, 0.960] & 0.960 [0.958, 0.961] & 0.940 [0.937, 0.944] \\
 & DP-SW-Probit ($\alpha=10$)   & 0.944 [0.936, 0.952] & 0.955 [0.950, 0.959] & 0.955 [0.950, 0.960] & 0.959 [0.955, 0.961] & 0.938 [0.934, 0.943] \\
 & DP-SW-Probit ($\alpha=100$)  & 0.949 [0.944, 0.953] & 0.955 [0.950, 0.957] & 0.958 [0.955, 0.961] & 0.952 [0.945, 0.958] & 0.940 [0.936, 0.945] \\
 & DP-SW-Probit ($\alpha=1000$) & 0.956 [0.953, 0.959] & 0.958 [0.952, 0.961] & 0.959 [0.954, 0.962] & 0.956 [0.940, 0.962] & 0.939 [0.933, 0.945] \\
\addlinespace
\multirow{7}{*}{7}
 & DP Probit                    & 0.883 [0.867, 0.902] & 0.912 [0.890, 0.931] & 0.903 [0.878, 0.914] & 0.917 [0.895, 0.930] & 0.901 [0.895, 0.908] \\
 & DP-SW-Probit ($\alpha=1$)    & 0.917 [0.913, 0.923] & 0.940 [0.935, 0.945] & 0.936 [0.928, 0.946] & 0.942 [0.937, 0.945] & 0.909 [0.902, 0.920] \\
 & DP-SW-Probit ($\alpha=10$)   & 0.911 [0.905, 0.920] & 0.918 [0.909, 0.928] & 0.922 [0.909, 0.938] & 0.941 [0.937, 0.944] & 0.915 [0.912, 0.920] \\
 & DP-SW-Probit ($\alpha=100$)  & 0.911 [0.877, 0.926] & 0.927 [0.901, 0.937] & 0.931 [0.908, 0.944] & 0.928 [0.920, 0.937] & 0.911 [0.897, 0.922] \\
 & DP-SW-Probit ($\alpha=1000$) & 0.936 [0.928, 0.941] & 0.929 [0.901, 0.944] & 0.929 [0.893, 0.945] & 0.933 [0.893, 0.952] & 0.902 [0.896, 0.911] \\
\bottomrule
\end{tabular}
}
\end{table}

\begin{table}[!t]
\centering
\footnotesize
\caption{Temporal hold-out, out-of-sample AUC (\#1(b) in Table~\ref{table:comparison}), same as Table~\ref{tab:in_auc} by forecasting forward into the withheld final $n_h=10$ tracked days $\mathcal{A}_i^{\mathrm{te}}$ of each patient at horizons $\tau \in \{1,3,7\}$.}
\label{tab:out_auc}
\scalebox{0.8}{
\begin{tabular}{llccccc}
\toprule
$\tau$ & Method & s1 & s2 & s3 & s4 & s5 \\
\midrule
\multirow{7}{*}{1}
 & DP Probit                    & 0.920 [0.911, 0.928] & 0.915 [0.897, 0.930] & 0.907 [0.893, 0.918] & 0.908 [0.886, 0.920] & 0.873 [0.847, 0.894] \\
 & DP-SW-Probit ($\alpha=1$)    & 0.930 [0.922, 0.941] & 0.934 [0.920, 0.943] & 0.914 [0.896, 0.932] & 0.939 [0.926, 0.948] & 0.906 [0.890, 0.922] \\
 & DP-SW-Probit ($\alpha=10$)   & 0.928 [0.916, 0.944] & 0.938 [0.931, 0.944] & 0.927 [0.912, 0.938] & 0.935 [0.922, 0.941] & 0.915 [0.905, 0.930] \\
 & DP-SW-Probit ($\alpha=100$)  & 0.934 [0.925, 0.942] & 0.929 [0.920, 0.939] & 0.932 [0.916, 0.939] & 0.934 [0.918, 0.944] & 0.917 [0.905, 0.928] \\
 & DP-SW-Probit ($\alpha=1000$) & 0.932 [0.923, 0.941] & 0.924 [0.904, 0.937] & 0.921 [0.906, 0.934] & 0.924 [0.904, 0.940] & 0.908 [0.888, 0.923] \\
\addlinespace
\multirow{7}{*}{3}
 & DP Probit                    & 0.899 [0.889, 0.908] & 0.896 [0.879, 0.913] & 0.892 [0.878, 0.906] & 0.892 [0.868, 0.905] & 0.856 [0.833, 0.874] \\
 & DP-SW-Probit ($\alpha=1$)    & 0.916 [0.908, 0.926] & 0.917 [0.903, 0.928] & 0.895 [0.879, 0.913] & 0.925 [0.911, 0.936] & 0.891 [0.874, 0.909] \\
 & DP-SW-Probit ($\alpha=10$)   & 0.913 [0.901, 0.931] & 0.921 [0.913, 0.929] & 0.910 [0.893, 0.922] & 0.919 [0.906, 0.928] & 0.895 [0.884, 0.913] \\
 & DP-SW-Probit ($\alpha=100$)  & 0.923 [0.912, 0.933] & 0.916 [0.905, 0.929] & 0.919 [0.900, 0.928] & 0.921 [0.903, 0.934] & 0.896 [0.881, 0.914] \\
 & DP-SW-Probit ($\alpha=1000$) & 0.914 [0.902, 0.927] & 0.903 [0.878, 0.919] & 0.902 [0.885, 0.918] & 0.902 [0.881, 0.920] & 0.885 [0.865, 0.902] \\
\addlinespace
\multirow{7}{*}{7}
 & DP Probit                    & 0.857 [0.845, 0.869] & 0.861 [0.837, 0.883] & 0.857 [0.840, 0.871] & 0.858 [0.830, 0.875] & 0.826 [0.804, 0.844] \\
 & DP-SW-Probit ($\alpha=1$)    & 0.892 [0.880, 0.902] & 0.887 [0.868, 0.898] & 0.868 [0.849, 0.890] & 0.895 [0.878, 0.909] & 0.866 [0.847, 0.887] \\
 & DP-SW-Probit ($\alpha=10$)   & 0.891 [0.881, 0.904] & 0.883 [0.871, 0.891] & 0.871 [0.852, 0.887] & 0.891 [0.876, 0.900] & 0.868 [0.853, 0.887] \\
 & DP-SW-Probit ($\alpha=100$)  & 0.893 [0.873, 0.908] & 0.882 [0.863, 0.896] & 0.882 [0.855, 0.896] & 0.894 [0.874, 0.907] & 0.870 [0.850, 0.891] \\
 & DP-SW-Probit ($\alpha=1000$) & 0.883 [0.868, 0.897] & 0.866 [0.837, 0.886] & 0.867 [0.845, 0.885] & 0.870 [0.845, 0.893] & 0.856 [0.834, 0.872] \\
\bottomrule
\end{tabular}
}
\end{table}

\begin{table}[ht]
\centering
\footnotesize
\caption{Patient hold-out, in-sample AUC (2A in Table~\ref{table:comparison}), same as Table~\ref{tab:in_auc} for the training patients $\mathcal{M}^{\mathrm{tr}}$ whose full trajectories were used to fit the model.}
\label{tab:patients_in_auc}
\scalebox{0.8}{
\begin{tabular}{llccccc}
\toprule
$\tau$ & Method & s1 & s2 & s3 & s4 & s5 \\
\midrule
\multirow{7}{*}{1}
 & DP Probit                    & 0.952 [0.942, 0.957] & 0.945 [0.935, 0.952] & 0.958 [0.950, 0.962] & 0.953 [0.942, 0.959] & 0.949 [0.939, 0.956] \\
 & DP-SW-Probit ($\alpha=1$)    & 0.970 [0.968, 0.971] & 0.970 [0.968, 0.971] & 0.971 [0.968, 0.972] & 0.970 [0.968, 0.972] & 0.970 [0.967, 0.972] \\
 & DP-SW-Probit ($\alpha=10$)   & 0.970 [0.968, 0.971] & 0.970 [0.969, 0.971] & 0.971 [0.969, 0.972] & 0.969 [0.966, 0.970] & 0.971 [0.968, 0.972] \\
 & DP-SW-Probit ($\alpha=100$)  & 0.968 [0.966, 0.969] & 0.968 [0.967, 0.969] & 0.971 [0.970, 0.971] & 0.968 [0.965, 0.969] & 0.970 [0.968, 0.971] \\
 & DP-SW-Probit ($\alpha=1000$) & 0.966 [0.964, 0.967] & 0.965 [0.964, 0.966] & 0.969 [0.967, 0.970] & 0.968 [0.966, 0.968] & 0.971 [0.968, 0.972] \\
\addlinespace
\multirow{7}{*}{3}
 & DP Probit                    & 0.944 [0.933, 0.950] & 0.931 [0.916, 0.940] & 0.946 [0.935, 0.952] & 0.938 [0.923, 0.947] & 0.932 [0.920, 0.942] \\
 & DP-SW-Probit ($\alpha=1$)    & 0.963 [0.960, 0.965] & 0.963 [0.960, 0.964] & 0.965 [0.961, 0.966] & 0.965 [0.961, 0.967] & 0.962 [0.957, 0.965] \\
 & DP-SW-Probit ($\alpha=10$)   & 0.961 [0.957, 0.964] & 0.965 [0.963, 0.966] & 0.965 [0.963, 0.967] & 0.961 [0.957, 0.963] & 0.964 [0.960, 0.967] \\
 & DP-SW-Probit ($\alpha=100$)  & 0.961 [0.957, 0.963] & 0.961 [0.959, 0.963] & 0.964 [0.962, 0.966] & 0.959 [0.955, 0.962] & 0.963 [0.961, 0.965] \\
 & DP-SW-Probit ($\alpha=1000$) & 0.951 [0.930, 0.960] & 0.954 [0.943, 0.958] & 0.958 [0.937, 0.964] & 0.956 [0.940, 0.961] & 0.961 [0.951, 0.965] \\
\addlinespace
\multirow{7}{*}{7}
 & DP Probit                    & 0.896 [0.887, 0.902] & 0.912 [0.897, 0.920] & 0.921 [0.908, 0.929] & 0.916 [0.887, 0.925] & 0.912 [0.900, 0.922] \\
 & DP-SW-Probit ($\alpha=1$)    & 0.944 [0.940, 0.949] & 0.943 [0.935, 0.949] & 0.945 [0.941, 0.949] & 0.949 [0.943, 0.953] & 0.935 [0.926, 0.942] \\
 & DP-SW-Probit ($\alpha=10$)   & 0.950 [0.942, 0.953] & 0.953 [0.949, 0.956] & 0.947 [0.941, 0.951] & 0.941 [0.933, 0.945] & 0.937 [0.926, 0.948] \\
 & DP-SW-Probit ($\alpha=100$)  & 0.946 [0.935, 0.951] & 0.949 [0.943, 0.953] & 0.950 [0.941, 0.955] & 0.946 [0.938, 0.950] & 0.950 [0.943, 0.953] \\
 & DP-SW-Probit ($\alpha=1000$) & 0.922 [0.850, 0.946] & 0.928 [0.902, 0.943] & 0.922 [0.887, 0.949] & 0.919 [0.871, 0.945] & 0.948 [0.927, 0.956] \\
\bottomrule
\end{tabular}
}
\end{table}

\begin{table}[!t]
\centering
\footnotesize
\caption{Patient hold-out, out-of-sample AUC (2B(b) in Table~\ref{table:comparison}), same as Table~\ref{tab:in_auc} for the target segments $\mathcal{A}_i^{\mathrm{te}}$ of held-out patients $\mathcal{M}^{\mathrm{te}}$, who contributed no data to model fitting; cluster assignment uses only each patient's pre-split history.}
\label{tab:patients_out_auc}
\scalebox{0.8}{
\begin{tabular}{llccccc}
\toprule
$\tau$ & Method & s1 & s2 & s3 & s4 & s5 \\
\midrule
\multirow{7}{*}{1}
 & DP Probit                    & 0.850 [0.764, 0.905] & 0.788 [0.645, 0.910] & 0.750 [0.595, 0.877] & 0.715 [0.586, 0.847] & 0.801 [0.687, 0.883] \\
 & DP-SW-Probit ($\alpha=1$)    & 0.810 [0.714, 0.880] & 0.810 [0.651, 0.869] & 0.784 [0.653, 0.883] & 0.767 [0.671, 0.871] & 0.744 [0.637, 0.842] \\
 & DP-SW-Probit ($\alpha=10$)   & 0.836 [0.751, 0.899] & 0.795 [0.700, 0.870] & 0.816 [0.673, 0.897] & 0.785 [0.628, 0.898] & 0.805 [0.654, 0.884] \\
 & DP-SW-Probit ($\alpha=100$)  & 0.855 [0.748, 0.905] & 0.858 [0.805, 0.904] & 0.819 [0.710, 0.888] & 0.846 [0.771, 0.902] & 0.831 [0.726, 0.895] \\
 & DP-SW-Probit ($\alpha=1000$) & 0.861 [0.822, 0.898] & 0.892 [0.863, 0.916] & 0.864 [0.815, 0.908] & 0.890 [0.779, 0.931] & 0.809 [0.647, 0.871] \\
\addlinespace
\multirow{7}{*}{3}
 & DP Probit                    & 0.803 [0.681, 0.882] & 0.738 [0.591, 0.883] & 0.701 [0.526, 0.849] & 0.654 [0.520, 0.807] & 0.747 [0.612, 0.858] \\
 & DP-SW-Probit ($\alpha=1$)    & 0.747 [0.623, 0.838] & 0.781 [0.619, 0.852] & 0.715 [0.555, 0.854] & 0.682 [0.571, 0.814] & 0.669 [0.552, 0.782] \\
 & DP-SW-Probit ($\alpha=10$)   & 0.761 [0.635, 0.857] & 0.714 [0.581, 0.824] & 0.736 [0.566, 0.864] & 0.741 [0.568, 0.876] & 0.768 [0.585, 0.871] \\
 & DP-SW-Probit ($\alpha=100$)  & 0.806 [0.685, 0.876] & 0.831 [0.748, 0.887] & 0.768 [0.612, 0.862] & 0.806 [0.713, 0.886] & 0.795 [0.657, 0.878] \\
 & DP-SW-Probit ($\alpha=1000$) & 0.839 [0.788, 0.880] & 0.871 [0.831, 0.900] & 0.840 [0.769, 0.895] & 0.861 [0.734, 0.911] & 0.770 [0.594, 0.843] \\
\addlinespace
\multirow{7}{*}{7}
 & DP Probit                    & 0.739 [0.586, 0.842] & 0.680 [0.526, 0.843] & 0.624 [0.416, 0.799] & 0.563 [0.413, 0.722] & 0.700 [0.546, 0.828] \\
 & DP-SW-Probit ($\alpha=1$)    & 0.681 [0.540, 0.805] & 0.732 [0.568, 0.822] & 0.636 [0.450, 0.802] & 0.597 [0.471, 0.749] & 0.602 [0.471, 0.731] \\
 & DP-SW-Probit ($\alpha=10$)   & 0.683 [0.549, 0.799] & 0.642 [0.491, 0.778] & 0.674 [0.481, 0.834] & 0.701 [0.512, 0.856] & 0.724 [0.539, 0.847] \\
 & DP-SW-Probit ($\alpha=100$)  & 0.745 [0.581, 0.831] & 0.783 [0.677, 0.854] & 0.724 [0.533, 0.834] & 0.748 [0.639, 0.855] & 0.760 [0.607, 0.860] \\
 & DP-SW-Probit ($\alpha=1000$) & 0.808 [0.727, 0.860] & 0.841 [0.792, 0.882] & 0.781 [0.697, 0.858] & 0.841 [0.696, 0.901] & 0.745 [0.572, 0.828] \\
\bottomrule
\end{tabular}
}
\end{table}

\paragraph{Out-of-sample predictions}
Table~\ref{tab:out_auc} reports out-of-sample AUC under the \underline{temporal
hold-out} scheme (\#1(b)), evaluated on the final $n_h=10$ tracked days withheld
from model fitting for every patient. As expected, out-of-sample AUC
is uniformly lower than the corresponding in-sample AUC in
Table~\ref{tab:in_auc} since prediction here compounds forecast error
across the horizon $\tau$ using only pre-split treatment
history. Nonetheless, DP-SW-Probit continues to outperform DP-Probit
across every scenario $s_1$--$s_5$ and every horizon $\tau \in
\{1,3,7\}$. The margin widens as $\tau$ grows: at $\tau=1$, $s_5$,
DP-SW-Probit reaches 0.917 versus 0.873 for DP-Probit (a gap of
0.044), while at $\tau=7$ the best DP-SW-Probit configuration reaches
0.894 versus 0.826 (a gap of 0.068). The effect of $\alpha$ on
out-of-sample AUC here also shows no consistent monotonic trend, with
performance often peaking at intermediate values ($\alpha \in \{10,
100\}$) rather than at the extremes. The trajectory similarity loss
therefore appears to trade a small amount of in-sample fit for
improved generalization to unseen days, without requiring aggressive
weighting. As in the in-sample tables, AUC decreases and credible
intervals widen with $\tau$, reflecting the compounding uncertainty of
autoregressive forecasting. 

Table~\ref{tab:patients_out_auc} reports out-of-sample AUC under the
\underline{patient hold-out} scheme (\#2B(b)). The target segment of each held-out patient
is predicted after assigning that patient to a cluster using only
their pre-split history, so no information from these patients enters
model fitting at all. Compared to the temporal hold-out, AUC drops
substantially for both methods (e.g., DP-Probit falls from 0.920 to
0.850 at $\tau=1$, $s_1$) and credible intervals widen considerably
(e.g., [0.764, 0.905] versus [0.911, 0.928]). This phenomenon reflects
the smaller test set and the added difficulty of generalizing to
patients who contributed no training data. In contrast to all three
preceding tables, larger values of $\alpha$ now yield clear and
often substantial improvements. In particular, for $s_2$ at $\tau=1$,
AUC rises from 0.788 (DP-Probit) and 0.810 ($\alpha=1$) to 0.892 at
$\alpha=1000$, and this broadly increasing pattern in $\alpha$
holds for most scenario--horizon combinations. This is consistent with
the role of the trajectory similarity loss in cluster assignment for
new patients (Section~ \ref{subsec:prediction_and_clstering}). A
held-out patient's label $S_{i^*}$ is chosen from only a short
observed history, so up-weighting the SW-distance term makes
assignment rely more on matching to the typical-patient trajectories
$\thb_k^*$. This matters because the regression likelihood, estimated
from limited within-patient data, is comparatively noisy for a patient
the model has never seen. At low $\alpha$, DP-SW-Probit can in fact
underperform DP-Probit (e.g., $s_1$, $\tau=1$: 0.810 versus 0.850),
indicating that some minimal weight on trajectory similarity is
necessary before its benefit to cluster assignment is realized. As in
the other tables, AUC degrades with $\tau$ across all methods and
scenarios. This degradation is steeper than under the temporal
hold-out, consistent with forecast error compounding on top of an
already-uncertain cluster assignment for a never-before-seen patient. 

Taken together, the four tables suggest a consistent story about the role of $\alpha$. When a patient's own history is available for fitting (temporal hold-out, both in- and out-of-sample, and patient-hold-out in-sample), performance is only mildly sensitive to $\alpha$ with excessively large values occasionally destabilizing the regression fit at longer horizons. When no history at all is available for fitting and cluster membership must be inferred from a short observed segment alone (patient-hold-out out-of-sample), larger $\alpha$ substantially improves predictive accuracy by shifting the clustering rule toward trajectory similarity and away from a regression likelihood that is unreliable for an unseen patient. This points to $\alpha$ as controlling a genuine  trade-off between within-patient regression fit and between-patient trajectory generalization.

\bibliography{example_paper}
\bibliographystyle{abbrv}

\end{document}

%% file: math_commands.tex
\usepackage{amsmath,amsfonts,bm}

\def\eqref#1{equation~\ref{#1}}
\def\1{\bm{1}}

\def\rb{{\textnormal{b}}}

\DeclareMathAlphabet{\mathsfit}{\encodingdefault}{\sfdefault}{m}{sl}
\SetMathAlphabet{\mathsfit}{bold}{\encodingdefault}{\sfdefault}{bx}{n}

%% file: example_paper.bib
@book{diggle2002analysis,
  title={Analysis of longitudinal data},
  author={Diggle, Peter},
  year={2002},
  publisher={Oxford university press}
}

@article{dahl2022search,
  title={Search algorithms and loss functions for Bayesian clustering},
  author={Dahl, David B and Johnson, Devin J and M{\"u}ller, Peter},
  journal={Journal of Computational and Graphical Statistics},
  volume={31},
  number={4},
  pages={1189--1201},
  year={2022},
  publisher={Taylor \& Francis}
}

@article{chiang2020individualizing,
  title={Individualizing the definition of seizure clusters based on temporal clustering analysis},
  author={Chiang, Sharon and Haut, Sheryl R and Ferastraoaru, Victor and Rao, Vikram R and Baud, Maxime O and Theodore, William H and Moss, Robert and Goldenholz, Daniel M},
  journal={Epilepsy Research},
  volume={163},
  pages={106330},
  year={2020},
  publisher={Elsevier}
}

@article{fisher2012seizure,
  title={Seizure diaries for clinical research and practice: limitations and future prospects},
  author={Fisher, Robert S and Blum, David E and DiVentura, Bree and Vannest, Jennifer and Hixson, John D and Moss, Robert and Herman, Susan T and Fureman, Brandy E and French, Jacqueline A},
  journal={Epilepsy \& Behavior},
  volume={24},
  number={3},
  pages={304--310},
  year={2012},
  publisher={Elsevier}
}

@inproceedings{berndt1994using,
  title={Using dynamic time warping to find patterns in time series},
  author={Berndt, Donald J and Clifford, James},
  booktitle={Proceedings of the 3rd International Conference on Knowledge Discovery and Data Mining},
  pages={359--370},
  year={1994}
}

@article{hoppe2007epilepsy,
  title={Epilepsy: accuracy of patient seizure counts},
  author={Hoppe, Christian and Poepel, Annkathrin and Elger, Christian E},
  journal={Archives of Neurology},
  volume={64},
  number={11},
  pages={1595--1599},
  year={2007},
  publisher={American Medical Association}
}

@article{miller2024long,
  title={Long-term seizure diary tracking habits in clinical studies: evidence from the Human Epilepsy Project},
  author={Miller, Kristen R and Barnard, Sarah and Juarez-Colunga, Elizabeth and French, Jacqueline A and Pellinen, Jacob and Human Epilepsy Project Investigators and others},
  journal={Epilepsy Research},
  volume={203},
  pages={107379},
  year={2024},
  publisher={Elsevier}
}

@article{haut2006seizure,
  title={Seizure clustering},
  author={Haut, Sheryl R},
  journal={Epilepsy \& Behavior},
  volume={8},
  number={1},
  pages={50--55},
  year={2006},
  publisher={Elsevier}
}

@article{pellinen2020focal,
  title={Focal nonmotor versus motor seizures: the impact on diagnostic delay in focal epilepsy},
  author={Pellinen, Jacob and Tafuro, Erica and Yang, Annie and Price, Dana and Friedman, Daniel and Holmes, Manisha and Barnard, Sarah and Detyniecki, Kamil and Hegde, Manu and Hixson, John and others},
  journal={Epilepsia},
  volume={61},
  number={12},
  pages={2643--2652},
  year={2020},
  publisher={Wiley Online Library}
}

@article{jafarpour2019seizure,
  title={Seizure cluster: definition, prevalence, consequences, and management},
  author={Jafarpour, Saba and Hirsch, Lawrence J and Ga{\'\i}nza-Lein, Marina and Kellinghaus, Christoph and Detyniecki, Kamil},
  journal={Seizure},
  volume={68},
  pages={9--15},
  year={2019},
  publisher={Elsevier}
}

@article{bauman2021seizure,
  title={Seizure clusters: morbidity and mortality},
  author={Bauman, Kristie and Devinsky, Orrin},
  journal={Frontiers in Neurology},
  volume={12},
  pages={636045},
  year={2021},
  publisher={Frontiers Media SA}
}

@article{mcinnes2018umap,
  title={UMAP: Uniform Manifold Approximation and Projection},
  author={McInnes, Leland and Healy, John and Saul, Nathaniel and Gro{\ss}berger, Lukas},
  journal={Journal of Open Source Software},
  volume={3},
  number={29},
  year={2018}
}

@article{chib1998analysis,
  title={Analysis of multivariate {P}robit models},
  author={Chib, Siddhartha and Greenberg, Edward},
  journal={Biometrika},
  volume={85},
  number={2},
  pages={347--361},
  year={1998},
  publisher={Oxford University Press}
}

@article{amato2025mmm,
  title={MMM: Clustering Multivariate Longitudinal Mixed-type Data},
  author={Amato, Francesco and Jacques, Julien},
  journal={arXiv preprint arXiv:2509.12166},
  year={2025}
}

@article{cantoni2025borrowing,
  title={Borrowing strength between unaligned binary time-series via {B}ayesian nonparametric rescaling of Unified Skewed {N}ormal priors},
  author={Cantoni, Beatrice and Poli, Giovanni and Juarez-Colunga, Elizabeth and M{\~A}{\v{z}}ller, Peter},
  journal={arXiv preprint arXiv:2505.06491},
  year={2025}
}

@article{kent2018personalized,
  title={Personalized evidence based medicine: predictive approaches to heterogeneous treatment effects},
  author={Kent, David M and Steyerberg, Ewout and Van Klaveren, David},
  journal={Bmj},
  volume={363},
  year={2018},
  publisher={British Medical Journal Publishing Group}
}

@article{fisher2010tracking,
  title={Tracking epilepsy with an electronic diary.},
  author={Fisher, Robert S},
  journal={Acta Paediatrica},
  volume={99},
  number={4},
  year={2010}
}

@article{kanaster2026mixed,
  title={A Mixed Self-Exciting Process to Model Epileptic Seizures},
  author={Kanaster, Karen and Silva, Giovani L and Mueller, Peter and Pellinen, Jacob and Juarez-Colunga, Elizabeth},
  journal={arXiv preprint arXiv:2605.22038},
  year={2026}
}

@incollection{varadhan2013estimation,
  title={Estimation and reporting of heterogeneity of treatment effects},
  author={Varadhan, Ravi and Seeger, John D},
  booktitle={Developing a Protocol for Observational Comparative Effectiveness Research: A User's Guide},
  year={2013},
  publisher={Agency for Healthcare Research and Quality (US)}
}

@article{nagin1999analyzing,
  title={Analyzing developmental trajectories: a semiparametric, group-based approach.},
  author={Nagin, Daniel S},
  journal={Psychological Methods},
  volume={4},
  number={2},
  pages={139},
  year={1999},
  publisher={American Psychological Association}
}

@book{fitzmaurice2012applied,
  title={Applied longitudinal analysis},
  author={Fitzmaurice, Garrett M and Laird, Nan M and Ware, James H},
  year={2012},
  publisher={John Wiley \& Sons}
}

@misc{French_HumanEpilepsyProject,
  author       = {French, Jacqueline A. and Kuzniecky, Ruben and Lowenstein, Daniel},
  title        = {Human Epilepsy Project},
  howpublished = {NYU Data Catalog},
  url          = {https://datacatalog.med.nyu.edu/dataset/10473},
  note         = {Dataset, UID 10473; accessed 2026-07-20},
  year = 2012
}

@inproceedings{rabin2012wasserstein,
  title={Wasserstein barycenter and its application to texture mixing},
  author={Rabin, Julien and Peyr{\'e}, Gabriel and Delon, Julie and Bernot, Marc},
  booktitle={Scale Space and Variational Methods in Computer Vision: Third International Conference, SSVM 2011, Ein-Gedi, Israel, May 29--June 2, 2011, Revised Selected Papers 3},
  pages={435--446},
  year={2012},
  organization={Springer}
}

@article{meilua2007comparing,
  title={Comparing clusterings—an information based distance},
  author={Meil{\u{a}}, Marina},
  journal={Journal of Multivariate Analysis},
  volume={98},
  number={5},
  pages={873--895},
  year={2007},
  publisher={Elsevier}
}

@article{wade2018bayesian,
  title={Bayesian Cluster Analysis: Point Estimation and Credible Balls (with Discussion)},
  author={Wade, Sara and Ghahramani, Zoubin},
  journal={Bayesian Analysis},
  volume={13},
  number={2},
  pages={559--626},
  year={2018}
}

@article{bonneel2015sliced,
  title={Sliced and {R}adon {W}asserstein Barycenters of Measures},
  author={Bonneel, Nicolas and Rabin, Julien and Peyr{\'e}, Gabriel and Pfister, Hanspeter},
  journal={Journal of Mathematical Imaging and Vision},
  volume={1},
  number={51},
  pages={22--45},
  year={2015}
}

@article{peyre2020computational,
  title={Computational optimal transport: With applications to data science},
  author={Peyr{\'e}, Gabriel and Cuturi, Marco and others},
  journal={Foundations and Trends in Machine Learning},
  volume={11},
  number={5-6},
  pages={355--607},
  year={2019},
  publisher={Now Publishers, Inc.}
}

@article{nguyen2023energy,
  title={Energy-Based Sliced {W}asserstein Distance},
  author={Nguyen, Khai and Ho, Nhat},
  journal={Advances in Neural Information Processing Systems},
  year={2023}
}

@inproceedings{
nguyen2024quasimonte,
title={Quasi-{M}onte {C}arlo for 3D Sliced {W}asserstein},
author={Khai Nguyen and Nicola Bariletto and Nhat Ho},
booktitle={The Twelfth International Conference on Learning Representations},
year={2024}
}

@article{bissiri2016general,
  title={A general framework for updating belief distributions},
  author={Bissiri, Pier Giovanni and Holmes, Chris C and Walker, Stephen G},
  journal={Journal of the Royal Statistical Society Series B: Statistical Methodology},
  volume={78},
  number={5},
  pages={1103--1130},
  year={2016},
  publisher={Oxford University Press}
}

@article{nguyen2025introduction,
  title={An Introduction to Sliced Optimal Transport: Foundations, Advances, Extensions, and Applications},
  author={Nguyen, Khai},
  journal={Foundations and Trends{\textregistered} in Computer Graphics and Vision},
  volume={17},
  number={3-4},
  pages={171--391},
  year={2025},
  publisher={Emerald Publishing Limited}
}

@article{neal2000markov,
  title={Markov chain sampling methods for {D}irichlet process mixture models},
  author={Neal, Radford M},
  journal={Journal of Computational and Graphical Statistics},
  volume={9},
  number={2},
  pages={249--265},
  year={2000},
  publisher={Taylor \& Francis}
}

@article{ferguson1973bayesian,
  title={A Bayesian analysis of some nonparametric problems},
  author={Ferguson, Thomas S},
  journal={The Annals of Statistics},
  pages={209--230},
  year={1973},
  publisher={JSTOR}
}
